\documentclass[10pt,journal,twoside,final]{IEEEtran}

\newcommand{\etal}{\textit{et al.}}

\usepackage{tablefootnote}
\usepackage{cite}
\usepackage{amsmath}
\usepackage{bigstrut}
\usepackage{graphicx, subfig}
\usepackage{multirow}

\usepackage{color}

\newcommand{\gx}[1]{{\color{black}#1}}

\begin{document}

\title{Towards Scalable and Channel-Robust Radio Frequency Fingerprint Identification for LoRa}


\author{Guanxiong~Shen,
	Junqing~Zhang,
	Alan~Marshall,~\IEEEmembership{Senior Member,~IEEE},
	and Joseph~Cavallaro,~\IEEEmembership{Fellow,~IEEE}	
	
	\thanks{Manuscript received xxx; revised xxx; accepted xxx. Date of publication xxx; date of current version xxx. The work was in part supported by UK Royal Society Research Grants under grant ID RGS\slash R1\slash 191241 and National Key Research and Development Program of China under grant ID 2020YFE0200600.
		The review of this paper was coordinated by xxx. 
		\textit{(Corresponding author: Junqing Zhang.)}}
	\thanks{G.~Shen, J.~Zhang and A.~Marshall are with the Department of Electrical Engineering and Electronics, University of Liverpool, Liverpool, L69 3GJ, United Kingdom. (email: Guanxiong.Shen@liverpool.ac.uk; junqing.zhang@liverpool.ac.uk; alan.marshall@liverpool.ac.uk)}
	\thanks{J.~Cavallaro is with the Department of Electrical and Computer Engineering, Rice University, Houston, USA. (email: cavallar@rice.edu)}
	\thanks{Color versions of one or more of the figures in this paper are available online at http://ieeexplore.ieee.org.}
	\thanks{Digital Object Identifier xxx}	
}

\maketitle

\begin{abstract}
Radio frequency fingerprint identification (RFFI) is a promising device authentication technique based on the transmitter hardware impairments. 
In this paper, we propose a scalable and robust RFFI framework achieved by deep learning powered radio frequency fingerprint (RFF) extractor. Specifically, we leverage the deep metric learning to train an RFF extractor, which has excellent generalization ability and can extract RFFs from previously unseen devices. Any devices can be enrolled via the pre-trained RFF extractor and the RFF database can be maintained efficiently for allowing devices to join and leave.
Wireless channel impacts the RFF extraction and is tackled by exploiting  channel independent feature and data augmentation.
We carried out extensive experimental evaluation involving 60 commercial off-the-shelf LoRa devices and a USRP N210 software defined radio platform. The results have successfully demonstrated that our framework can achieve excellent generalization abilities for device classification and rogue device detection as well as effective channel mitigation. 

\end{abstract}

\begin{IEEEkeywords}
Internet of things, device authentication,  radio frequency fingerprint identification, deep learning, LoRa
\end{IEEEkeywords}	



\section{Introduction}
With the rapid growth in the population of the Internet of things (IoT) devices, their security is becoming increasingly important. Device authentication is required to prevent the intrusion of rogue devices that may execute harmful processes or access to private data~\cite{yang2017survey}. 
It is usually achieved by cryptographic schemes such as the common challenge-response-based protocols. These schemes require a common key pre-shared which may be difficult for IoT devices because they are usually low-cost and power constrained~\cite{zhang2020new}. They also rely on software addresses such as MAC address, which can however be tampered with easily.

Lightweight and reliable device authentication is thus urgently required for ensuring IoT security. Radio frequency fingerprint identification (RFFI) is a promising non-cryptographic technique that relies on the device intrinsic hardware impairments of radio frequency (RF) components generated during the manufacturing process. 
The hardware characteristics of these components slightly deviate from their nominal values but the deviation is so small that the normal communication functionality is not affected.
These hardware features are unique and difficult to be tampered with, thus they can be treated as the fingerprint of IoT devices, in a similar manner as the biometrics of human beings.
The transmitter hardware impairments will distort the transmission waveform, from which the receiver can infer the transmitter identity by extracting the device-specific radio frequency fingerprints (RFFs). RFFI can thus identify transmitters on a per-packet basis. All the RFFI operations are completed at the receiver and there is no modification of the transmitter. Hence, it is particularly suitable for power-constrained and low-cost IoT devices. 

The existing RFFI work can be categorized into traditional RFF extractor-based and deep learning-based approaches. The former relies on the manually designed RFF extractor to obtain hardware features such as IQ imbalance~\cite{brik2008wireless,shi2011improved}, CFO~\cite{liu2019real,joo2020hold,hua2018accurate,shi2011improved,polak2015wireless,peng2018design}, signal transient part~\cite{danev2009transient,xing2020design}, power spectral density~\cite{wang2016wireless,danev2009physical}, power amplifier non-linearity~\cite{polak2011identifying}, beam pattern~\cite{balakrishnan2019physical}, etc. However, such schemes are highly dependent on the quality of the designed feature extraction algorithms and require a deep understanding on the adopted communication protocol. Some hardware characteristics are interrelated with each other, making it challenging to extract each feature individually. On the other hand,  deep learning-based approaches usually rely on a classification neural network to process raw signals and directly infer device identity without the need of feature engineering, which has attracted wide attention in recent years~\cite{al2020exposing,roy2019rfal,cekic2020robust,agadakos2020chameleons,shen2021radio,robyns2017physical,das2018deep,yu2019robust,reus2021classifying,jian2021radio,soltani2020rf,peng2019deep,he2020cooperative,zhang2021radio,liao2019multiuser,qian2021specific,gong2020unsupervised,rajendran2020injecting,al2021deeplora,piva2021tags}. 

\begin{figure}[!t]
	\centering
	\includegraphics[width=3in]{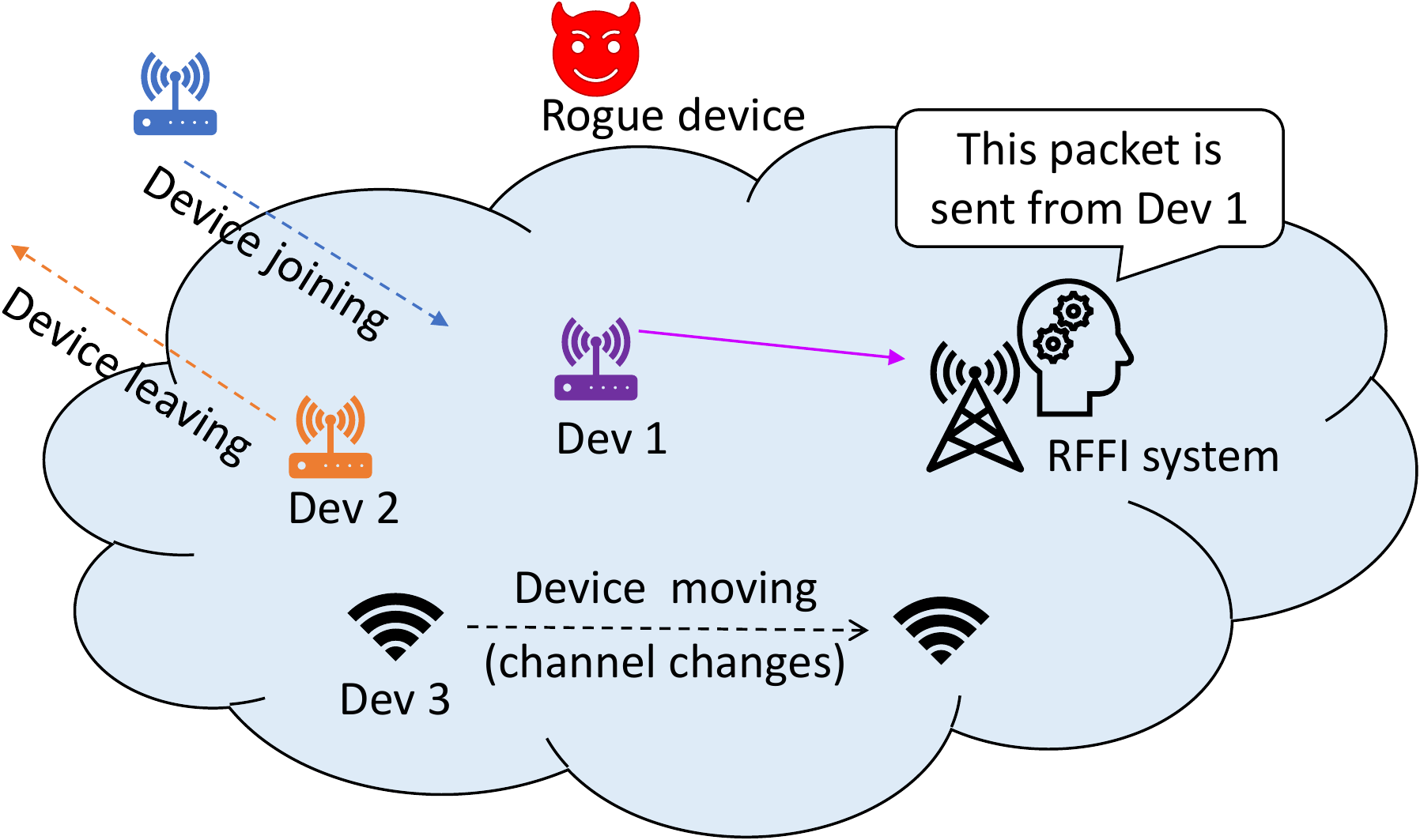}
	\caption{IoT networks secured by RFFI. Deep learning-based RFFI systems suffer from limited scalability and channel robustness.}
	\label{fig:rffi}
\end{figure}

While deep learning-based RFFI has excellent classification accuracy, it suffers from the limited scalability and channel robustness issues, as shown in Fig.~\ref{fig:rffi}. The number of neurons of the softmax layer in deep learning-based RFFI systems is unchangeable after training, which restricts the RFFI to a close-set problem. Hence, 
the scalability of such design is limited as only a fixed number of devices can be identified~\cite{hanna2020open,zhou2021robust}, which raises two concerns:
\begin{itemize}
	\item The deep learning model needs to be re-trained whenever legitimate devices join and leave the IoT network, which is time-consuming.
	\item Rogue devices cannot be predicted and their data is usually not available for training. During the classification, they will be classified as the legitimate devices with the most similar characteristics in the training categories~\cite{shen2021radio,das2018deep,zhang2021radio}, which is not acceptable.
\end{itemize}
Moreover, all the RFFI systems are subject to channel effects as the received signal is not only distorted by hardware impairments but also the wireless channel.


\gx{In our prior work, we explored the selection of  signal representation (IQ samples, FFT coeefficients and spectrogram), stability, and deep learning models for RFFI systems~\cite{shen2021jsac}. However, the above mentioned challenges, i.e., the scalability and channel robustness, are not well addressed. This paper aims to design a scalable and channel-agnostic RFFI system. 
To achieve this, our approach  introduces an enrollment stage and uses k-nearest neighbor (k-NN) to replace the softmax layer for recognition. The channel independent spectrogram is also constructed to mitigate channel effects.}
Extensive experiments were carried out with 60 commercial off-the-shelf LoRa devices of four models at various channel conditions to demonstrate the excellent performance of the proposed RFFI system.
Our contributions are highlighted as follows.
\begin{itemize}
	\item \gx{We design a scalable RFFI framework based on a deep metric learning-powered RFF extractor, which enables device joining and leaving without the need for re-training.} It maintains an RFF database by enrolling a new device using the pre-trained RFF extractor or deleting the record of a leaving device.
	\item We propose a channel robust RFFI protocol by constructing  the channel independent spectrogram and exploiting data augmentation. The channel independent spectrogram can mitigate the channel effect in the time-frequency domain while reserving the RFF of the LoRa signal. The data augmentation is carefully designed to represent real channel conditions with both multipath and Doppler shift.
	\item We conduct extensive experiments involving different LoRa devices, various channel conditions and antenna polarization. We experimentally demonstrate that the RFF extractor is able to extract features from devices that are not present during training, even they are from other manufacturers. The proposed channel independent spectrogram is shown to be effective in mitigating the channel effect and data augmentation can further increase the system robustness. Antenna polarization is found to affect the classification performance.	
\end{itemize}

\gx{This paper provides a general RFFI framework and studies LoRa as an example. 
}
LoRa devices are manufactured with low-cost components therefore have abundant hardware impairments, which are  suitable for RFFI. \gx{Moreover, the commercial LoRaWAN specification defines cryptography-based device authentication schemes. The root keys are assigned to the end-nodes during fabrication and secure storage of them is a huge challenge. The proposed RFFI can provide an alternative method to identify and authenticate LoRa devices.} 
%

The rest of the paper is organized as follows. Section~\ref{sec:system_overview} shows the system overview. Section~\ref{sec:lora_signal_processing} introduces the LoRa signal processing. Section~\ref{sec:channel_independent_spectrogram} presents the procedure of constructing the channel independent spectrogram. The training of RFF extractor is introduced in Section~\ref{sec:rff_extractor_training} while the enrollment and identification of RFFI systems is introduced in Section~\ref{sec:enrollment_identification}. Section~\ref{sec:experimental_evaluation} provides extensive evaluation results to show the system performance. Section~\ref{sec:related_work} and Section~\ref{sec:conclusion} introduces related work and concludes the paper, respectively.

\section{System Overview}\label{sec:system_overview}
\begin{figure}[!t]
	\centering
	\includegraphics[width=3.4in]{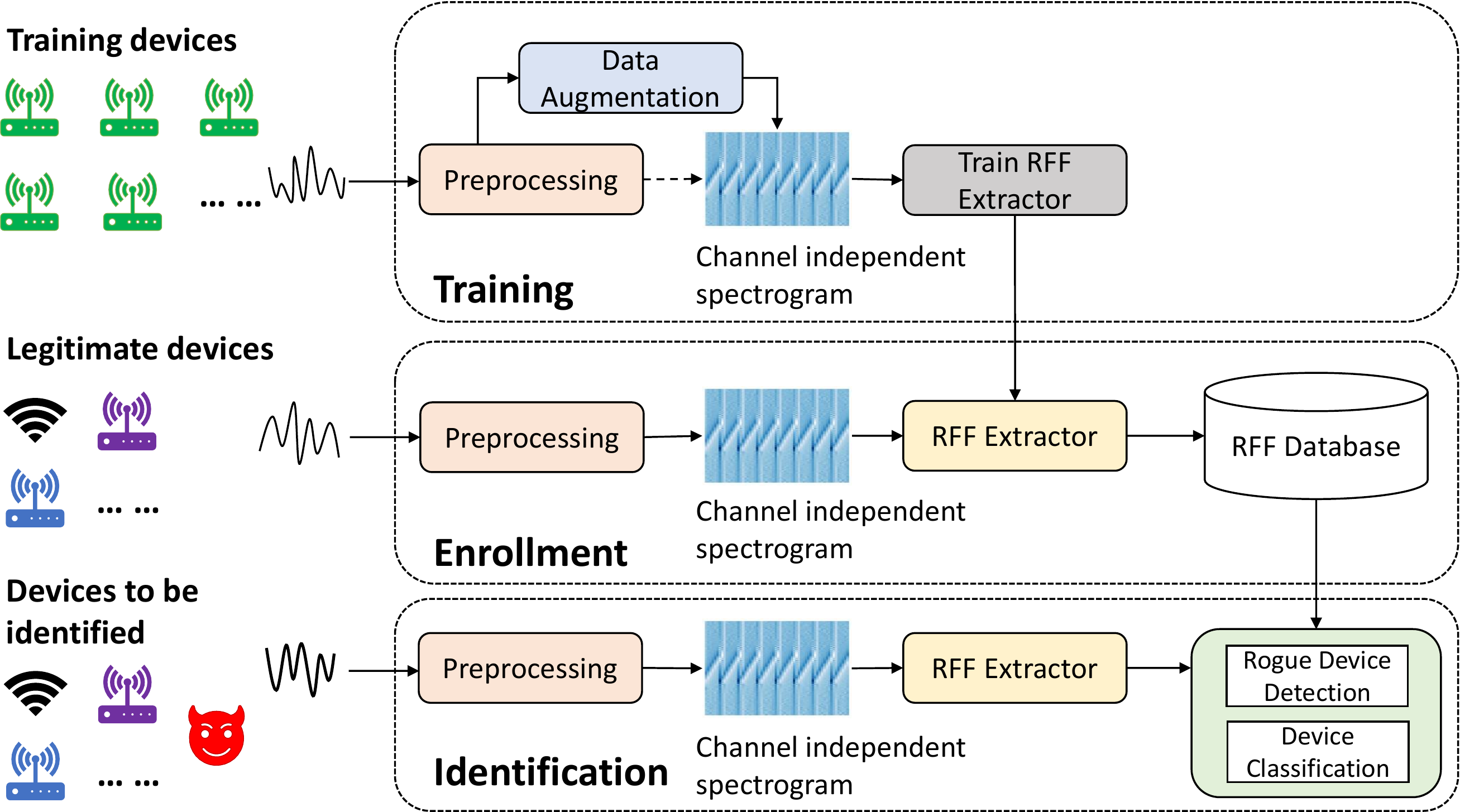}
	\caption{System diagram of the proposed RFFI system.}
	\label{fig:system_overview}
\end{figure}

The proposed system is based the deep learning-powered RFF extractor. As shown in Fig.~\ref{fig:system_overview}, it involves training, enrollment and identification stages.

\textbf{Training}: \gx{The training stage will leverage the outstanding feature extraction capability of deep learning to generate an RFF extractor rather than a classification neural network.} Specifically, a large number of training packets are collected from numerous training devices. The data augmentation is used to increase channel diversity. The training samples are then transformed to channel independent spectrograms to mitigate channel effects. Finally, we leverage the triplet loss in deep metric learning to train the extractor. The input of the RFF extractor is a 2-D channel independent spectrogram, and the output is a vector consists of 512 elements, which is the unique RFF of that device.

The training only needs to be done once as the trained RFF extractor is able to extract unique RFFs from out-of-library (unseen) devices. 
The training devices are not necessarily the same as the ones for enrollment and identification.
The training can be done by a third party that owns massive data collected from a huge number of devices to train an RFF extractor with excellent generalization capability.
	
\textbf{Enrollment}: The enrollment will obtain the RFFs of legitimate devices using the RFF extractor. These are the actual working devices in an IoT network, which are probably different from the training devices. These legitimate devices are required to send several packets to the RFFI system. Then RFFs can be obtained from the received packets using the trained RFF extractor and stored in a database.
	The RFFs of the newly joined devices will be added to the database and the RFFs of devices that leave the system will be deleted, which makes the system easily updated. 
	The enrollment should be carried out in a controlled environment to ensure the RFFs of rogue devices are not enrolled.
	
\textbf{Identification}: A complete identification system should consist of two parts, namely rogue device detection and device classification. Rogue device detection first determines whether the transmitter is legitimate (previously enrolled), and device classification further infers its label. Both of them are implemented by the k-NN algorithm.


	
	
	
	

\section{LoRa Signal Processing}\label{sec:lora_signal_processing}

\subsection{LoRa Signal}\label{sec:lora_channel}

LoRa is based on the chirp spread spectrum (CSS) technology which uses chirps for communication. The instantaneous frequency of the LoRa signal changes continuously over time, and a basic LoRa symbol (up-chirp) can be written as
\begin{equation} 
	u(t) = A e^{j2\pi(-\frac{bw}{2}+\frac{bw}{2 T} t)t} \quad (0 \leq t \leq T), 
\end{equation}
where $A$ and $bw$ denote amplitude and bandwidth, respectively. $T$ is the LoRa symbol duration. 
There are repeating up-chirps at the beginning of a LoRa packet called the preamble. The preamble is identical in every LoRa packet regardless of the device type.

\subsection{Signal Acquisition}
The baseband transmitted signal, $x(t)$, will undergo signal modulation and up-conversion via hardware components such as oscillator and power amplifier, etc.
These components will have their specific impairments and their overall effect is denoted as $f(\cdot)$. The signal then travels via the wireless channel and is captured by the receiver. The baseband received signal is given as
\begin{equation}\label{equ:received_signal}
	y(t) =  h(\tau,t)*f(x(t)) + n(t),
\end{equation}
where $h(\tau,t)$ is the time-varying channel impulse response,   $n(t)$ is the additive white Gaussian noise and $*$ denotes convolution operation.
The received signal, $y(t)$, is further converted to digital samples by an analog-digital-converter (ADC) with a sampling interval of $T_s$, denoted as $y[nT_s]$. It is simplified to $y[n]$ for easy notation. 

\subsection{Preprocessing}\label{sec:preprocessing}
The received signal needs to be pre-processed to meet the basic requirements of RFFI, including synchronization, carrier frequency offset (CFO) compensation and normalization. These algorithms are briefly described below and detailed descriptions can be found in our prior work~\cite{shen2021jsac}. 

\textbf{Synchronization}: Synchronization locates the starting point of the transmission. Inaccurate synchronization introduces a segment of channel noise, which will  affect the RFFI performance. 

\textbf{Preamble Extraction}: The deep learning model can learn the identity-related information such as the MAC address if the entire packet is used for RFFI. To prevent this, only the preamble part, $s'[n]$, is employed in the proposed system. 

\textbf{CFO Compensation}: CFO compensation is essentially required in RFFI for system stability. It has been proved that crystal oscillators are particularly sensitive to temperature changes~\cite{andrews2019extensions}, and the drift of oscillator frequency can seriously degrade the system performance~\cite{shen2021radio, cekic2020robust}.


\textbf{Normalization}: Normalization prevents the system from learning the received power that is not device-specific. The preamble part is normalized by dividing the root mean square (RMS) of it.
The preprocessed signal is denoted as $s[n]$.





\section{Channel Independent Spectrogram}\label{sec:channel_independent_spectrogram}
The received signal is not only distorted by the hardware impairments but also by the wireless channel. It is impossible to ensure the devices experience the same channel during enrollment and identification. We propose the channel independent spectrogram to mitigate the channel effect in the time-frequency domain while preserving the device-specific characteristics.

\subsection{Observation of Channel Effect}
LoRa supports three bandwidth options, namely 125~kHz, 250~kHz and 500~kHz. LoRa transmissions are sometimes assumed to experience flat fading where the wireless channel causes the same magnitude and phase changes to all the frequency components.

\begin{figure}[!t]
	\centering
	\subfloat[]{\includegraphics[width=1.6in]{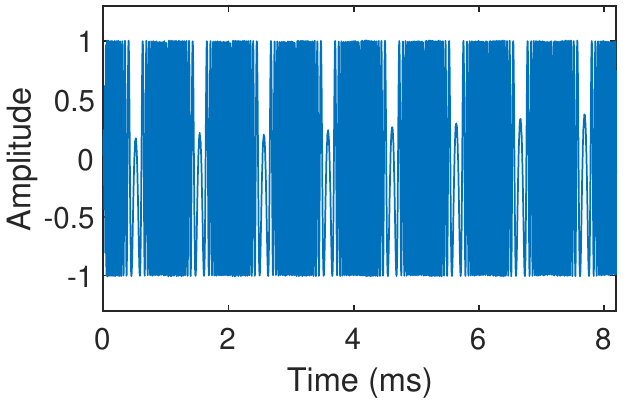}
		\label{fig:DUT6_LocA}}
	\subfloat[]{\includegraphics[width=1.6in]{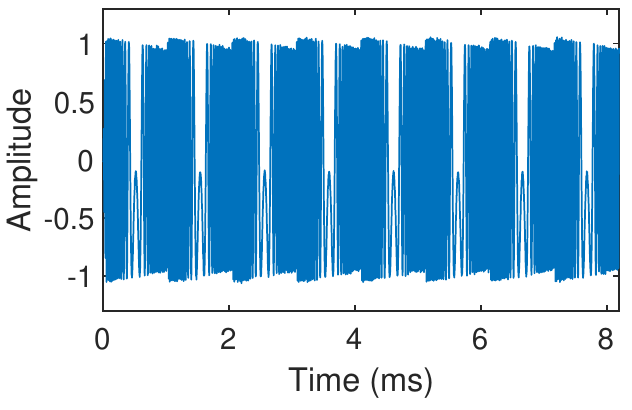}
		\label{fig:DUT6_LocF}}
	
	\subfloat[]{\includegraphics[width=1.6in]{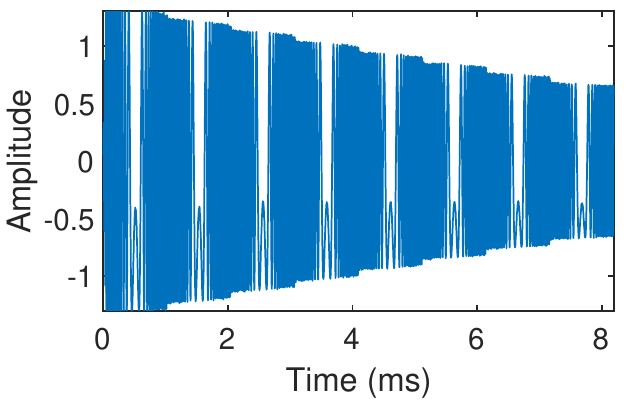}
		\label{fig:DUT1_walking_office}}
	\subfloat[]{\includegraphics[width=1.6in]{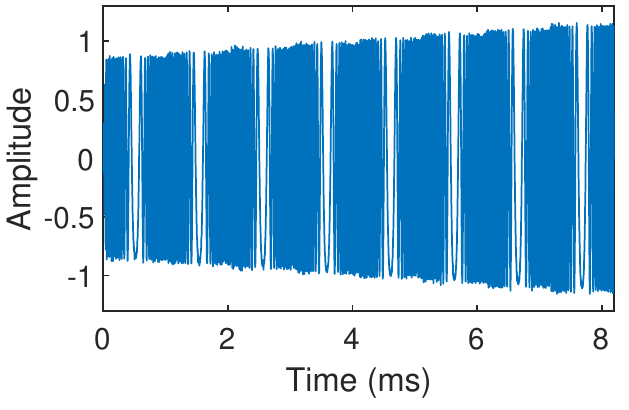}
		\label{fig:DUT1_walking_meeting_room}}
	\caption{Normalized received LoRa waveform. (a) LOS stationary scenario. (b) NLOS stationary scenario. (c) LOS mobile scenario. (d) NLOS mobile scenario.}
	\label{fig:collected_waveform}
\end{figure}

However, we experimentally found that the wireless channel significantly distorts the LoRa signal whose effect is evident from the received waveform. We collected LoRa packets in line-of-sight (LOS) stationary and non-line-of-sight (NLOS) stationary scenarios as well as LOS mobile and NLOS mobile scenarios. Detailed experimental setting can be found in Section~\ref{sec:experimental_evaluation}. The  preamble part (I-branch) is shown in Fig.~\ref{fig:collected_waveform}. It can be observed that the waveforms collected under various scenarios are different. We will later in Section~\ref{sec:data_augmentation} demonstrate that the sawtooth shapes are caused by the time dispersion (multipath effect) and the amplitude variation is due to channel changes (Doppler effect). 

\subsection{Short-time Fourier Transform}
\begin{figure}[!t]
	\centering
	\subfloat[]{\includegraphics[width=1.4in]{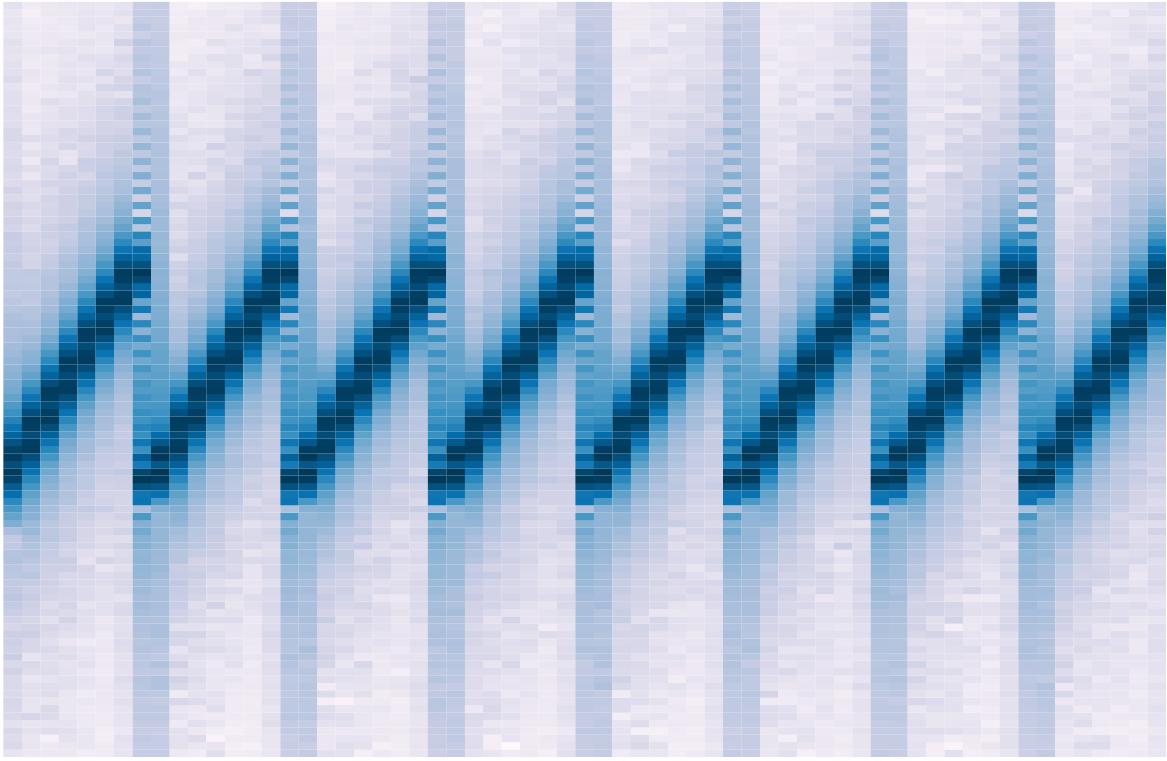}
		\label{fig:spectrogram}}
	\subfloat[]{\includegraphics[width=1.4in]{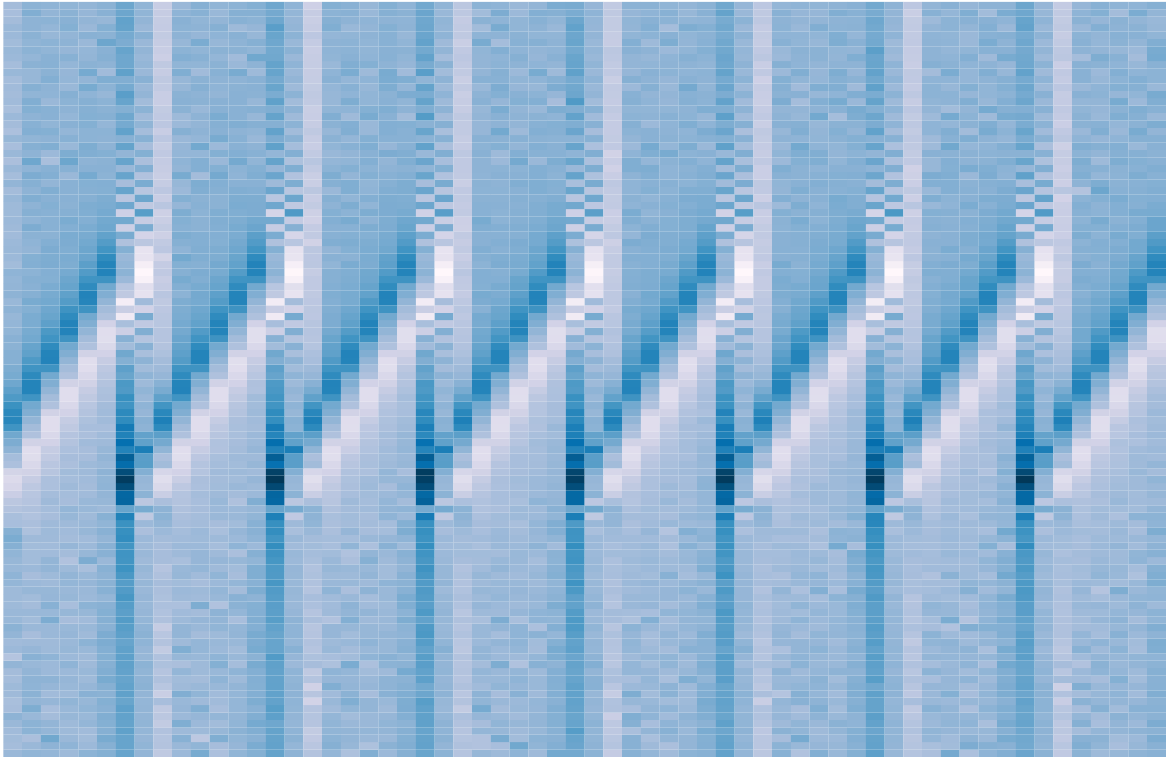}
		\label{fig:channel_independent_spectrogram}}
	\caption{(a) Spectrogram of LoRa preambles. (b) Channel independent spectrogram of LoRa preambles.}
	\label{fig:spectrogram_and_differential_spectrogram}
\end{figure}

The LoRa signal is usually analyzed in the time-frequency domain because of its non-stationary property. Short-time Fourier transform (STFT) is an efficient time-frequency analysis algorithm which can reveal the time-frequency features of the signal. The discrete STFT is mathematically written as
\begin{equation} \label{equ:stft}
	\begin{aligned}
		\mathbf{S}_{k,m} &= \sum_{n=0}^{N-1} s[n] w[n-mR]e^{-j2\pi \frac{k}{N} n}\\ 
		&\mbox{for}\ k = 1,2,..., N\ \mbox{and}\ m = 1,2,..., M,
	\end{aligned}
\end{equation}
where $\mathbf{S}_{k,m}$ is the element of the STFT complex matrix $\mathbf{S}$. $M$ is number of columns, $N$ is the length of window function $w[n]$, which is also the number of fast Fourier transform (FFT) points and rows, $R$ is the hop size. In our implementations, $N$ is 256, $R$ is 128, and $M$ is calculated as 63. The spectrogram in dB scale, $\widetilde{\mathbf{S}}$, is given as
\begin{equation}\label{equ:spectrogram}
	\widetilde{\mathbf{S}} = 10 \log_{10}(\left | \mathbf{S} \right |^{2}),
\end{equation}
where $\left |\cdot \right |$ returns the amplitude. 
The spectrogram of the LoRa preamble part is shown in Fig.~\ref{fig:spectrogram}.\\


\subsection{Constructing Channel Independent Spectrogram}
The spectrogram is affected by the wireless channel, which can be mitigated by dividing the adjacent columns.
STFT can be considered as spliting $s[n]$ into $M$ segments of $N$ samples, then performing FFT on each segment.  The overall effect of hardware distortion to the transmitted signal in the frequency domain is denoted as $F(\cdot)$.
Therefore, the STFT complex matrix $\mathbf{S}$ can be arranged as
\begin{equation}\label{equ:stft_matrix}
	\begin{bmatrix}
		H_{1,1} F(X_{1,1})&  H_{1,2} F(X_{1,2})& \cdots &  H_{1,M} F(X_{1,M})\\ 
		H_{2,1} F(X_{2,1})&  H_{2,2} F(X_{2,2})& \cdots &  H_{2,M}\ F(X_{2,M})\\ 
		\vdots &  \vdots &  \cdots &  \vdots\\ 
		H_{N,1} F(X_{N,1})&  H_{N,2} F(X_{N,2})& \cdots &  H_{N,M} F(X_{N,M}).
	\end{bmatrix}\\ \\ 
\end{equation}
where $[X_{1,m}, X_{2,m} \cdots X_{N,m}]^{T}$ denotes the ideal frequency spectrum of the $m$-th signal segment. Correspondingly, the column vector $[H_{1,m}, H_{2,m} \cdots H_{N,m}]^{T}$ represents the channel frequency response experienced by the $m$-th signal segment.
Therefore, (\ref{equ:stft_matrix}) can be simplified as
\begin{equation}\label{equ:stft_simple}
		\mathbf{S} = [\mathbf{H}_{1} \odot F(\mathbf{X}_{1})\quad \mathbf{H}_{2} \odot F(\mathbf{X}_{2})\   \cdots \  \mathbf{H}_{M} \odot F(\mathbf{X}_{M})],
\end{equation}
where $\odot$ denotes element-wise product. $\mathbf{X}_{m}$ represents the ideal frequency spectrum of the m-th signal segment and $\mathbf{H}_{m}$ represents corresponding channel frequency response.

In our experimental settings, the time gap between two adjacent signal segments is only 128~$\mu s$. It is reasonable to assume $\mathbf{H}_{m} \approx \mathbf{H}_{m+1}$ since the wireless channel will not change dramatically in such a short period. Based on this assumption, we can divide the m-th column by the (m+1)-th one so that the channel-related information can be  eliminated. The matrix after division is given as
\begin{equation}\label{equ:channel_independent} 
	\begin{aligned}
		\mathbf{Q} &= \Big[\frac{F(\mathbf{X}_{2})}{F(\mathbf{X}_{1})} \quad \frac{F(\mathbf{X}_{3})}{F(\mathbf{X}_{2})} \quad \cdots \quad \frac{F(\mathbf{X}_{M})}{F(\mathbf{X}_{M-1})}\Big].
	\end{aligned}
\end{equation}
Compared to~(\ref{equ:stft_simple}), the channel information is eliminated but device-specific hardware distortion $F(\cdot)$ are preserved. Similar to~(\ref{equ:spectrogram}), the magnitude of $\mathbf{Q}$ is expressed in dB scale
\begin{equation} 
		\widetilde{\mathbf{Q}} = 10 \log_{10}(\left | \mathbf{Q} \right |^{2}).
\end{equation}
$\widetilde{\mathbf{Q}}$ is the proposed channel independent spectrogram which is used as the input of RFF extractor. 
The channel independent spectrogram of LoRa preamble part is shown in Fig.~\ref{fig:channel_independent_spectrogram}. 




\section{RFF Extractor Training}\label{sec:rff_extractor_training}
The RFF extractor is the core module in the proposed RFFI system. It should extract channel independent and discriminative RFFs from the received signal and generalize well on previously unseen devices.



\subsection{Data Augmentation}\label{sec:data_augmentation}


Data augmentation is an efficient approach in deep learning to improve performance  and has been recently applied in the area of RFFI to increase its robustness to the wireless channel~\cite{soltani2020more, cekic2020robust, merchant2019enhanced}. Data augmentation can generate more training data to increase the performance of the RFF extractor and reduce the overhead for data collection.
It can also mitigate the channel effect by injecting various channel distortions into the training data so that the RFF extractor automatically learns how to deal with them.
 
In this paper, the channel effect for data augmentation includes both multipath and Doppler shift. The multipath is described by the power delay profile (PDP). The exponential PDP is selected and the discrete model is given as
\begin{equation}
	P(p) = \frac{1}{\tau_{d}}e^{-pT_{s}/\tau_{d}},\quad p = 0,1,\cdots,p_{max},
\end{equation}
where $\tau_{d}$ is the RMS delay spread and $p_{max}$ is the index of the last path. The PDP is further normalized.

This paper also considers Doppler shift, which is overlooked in previous work as the channel is assumed to be constant in a packet time~\cite{soltani2020more,cekic2020robust, merchant2019enhanced}. 
However, this assumption might not hold for some LoRa transmissions. Each LoRa preamble typically lasts about 1~$ms$ and the channel effect may not remain constant in mobile scenarios. Actually, the Doppler effect is observed from the received waveform  such as Fig.~\ref{fig:DUT1_walking_office} and Fig.~\ref{fig:DUT1_walking_meeting_room}. 
The Doppler effect can be characterized by the Doppler spectrum. This paper adopts the popular Jakes model whose spectrum is defined as
\begin{equation}
	S(f) = \frac{1}{\pi f_d \sqrt{1-(f/f_d)^{2}}},
\end{equation}
where $f_d$ is the maximum Doppler shift.



The data augmentation is carried out as follows.
\begin{enumerate}
\item The training data should be collected in a short-distance LOS stationary scenario so that it can be assumed as experiencing frequency-flat slow fading.
\item The number of training samples is increased by replication. It is doubled in our implementation.
\item The channel effect with multipath and Doppler effect is generated with the parameters randomly selected within specific ranges given in Table~\ref{tab:augmentation_parameters}.
\item The packet generated in step 1) passes through the channel from step 3). The artificial white Gaussian noise is added to the signal.
\end{enumerate}
The augmentation is completed by repeating the above process for all the training packets.
\begin{table}[!t]
	\centering
	\caption{Parameter of the Channel Simulator}
	\begin{tabular}{|l|r|}
		\hline
		Paramter & \multicolumn{1}{l|}{Range} \bigstrut\\
		\hline
		RMS delay spread $\tau_{d}$~(ns) & [5,300] \bigstrut\\
		\hline
		Maximum Doppler frequency $f_d$~(Hz) & [0,10] \bigstrut\\
		\hline
		Rician K-factor & [0,10] \bigstrut\\
		\hline
		SNR~(dB) & [20,80] \bigstrut\\
		\hline
	\end{tabular}%
	\label{tab:augmentation_parameters}%
\end{table}%

\begin{figure}[!t]
	\centering
	\subfloat[]{\includegraphics[width=1.6in]{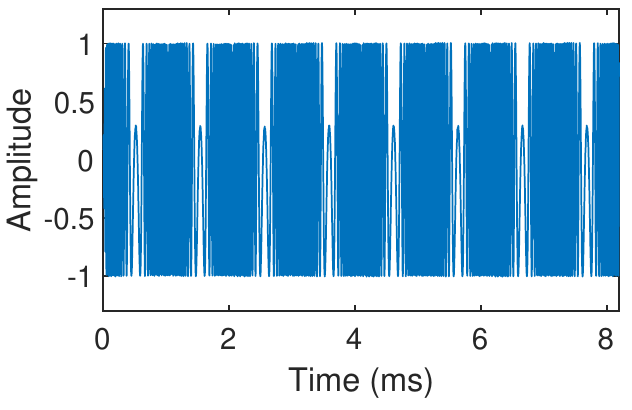}
		\label{fig:sim_awgn}}
	\subfloat[]{\includegraphics[width=1.6in]{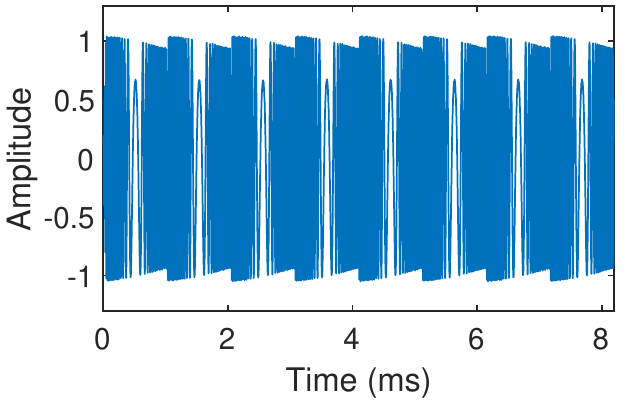}
		\label{fig:sim_multipath}}
	
	\subfloat[]{\includegraphics[width=1.6in]{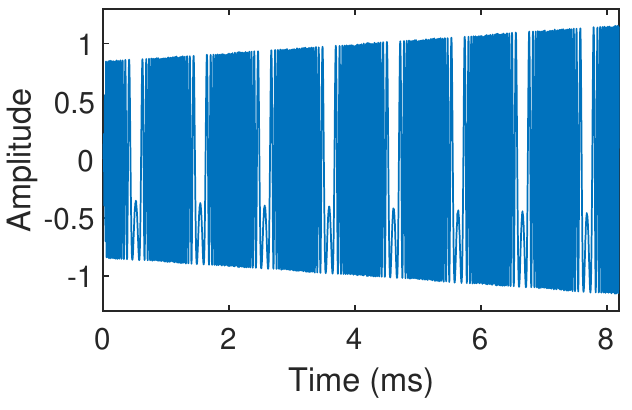}
		\label{fig:sim_doppler}}
	\subfloat[]{\includegraphics[width=1.6in]{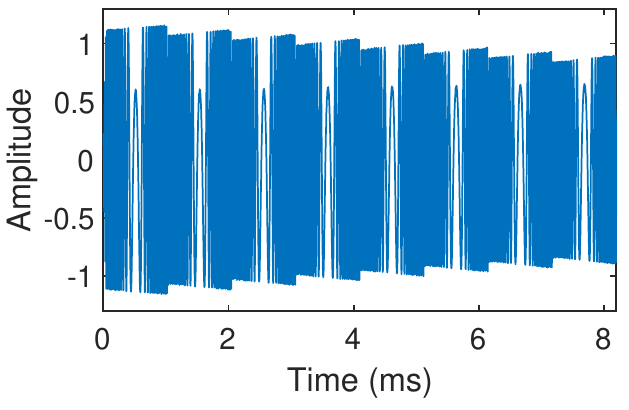}
		\label{fig:sim_both}}
	\caption{Augmented LoRa preambles. (a) Original waveform. (b) Strong multipath no Doppler effect ($\tau_d$ = 300~ns, $f_d$ = 0~Hz). (c) Strong Doppler effect with weak multipath ($\tau_d$ = 5~ns, $f_d$ = 10~Hz). (d) Both multipath and Doppler effects are strong ($\tau_d$ = 300~ns, $f_d$ = 10~Hz).}
	\label{fig:lora_channel_analysis}
\end{figure}

We show the waveform of three cases, namely strong multipath no Doppler effect, strong Doppler effect with weak multipath and both strong multipath and Doppler effects in Fig.~\ref{fig:sim_multipath}, Fig.~\ref{fig:sim_doppler} and Fig.~\ref{fig:sim_both}, respectively. It can be observed that the augmented waveforms match well with the real collected ones shown in Fig.~\ref{fig:collected_waveform}. We can also conclude that the wireless channel distorts the received LoRa signal significantly. The sawtooth shapes are caused by multipath and amplitude variation is caused by the Doppler effect.

\subsection{Model Architecture}
The channel independent spectrogram can be regarded as a 2-D image so the efficient image classification model convolutional neural network (CNN) is used to extract RFFs from it. The CNN in this paper is designed with reference to the well-known ResNet~\cite{he2016deep}. It is further optimized to be more lightweight and suitable for the size of channel independent spectrograms. 

\begin{figure}[!t]
	\centering
	\includegraphics[width=2.0in]{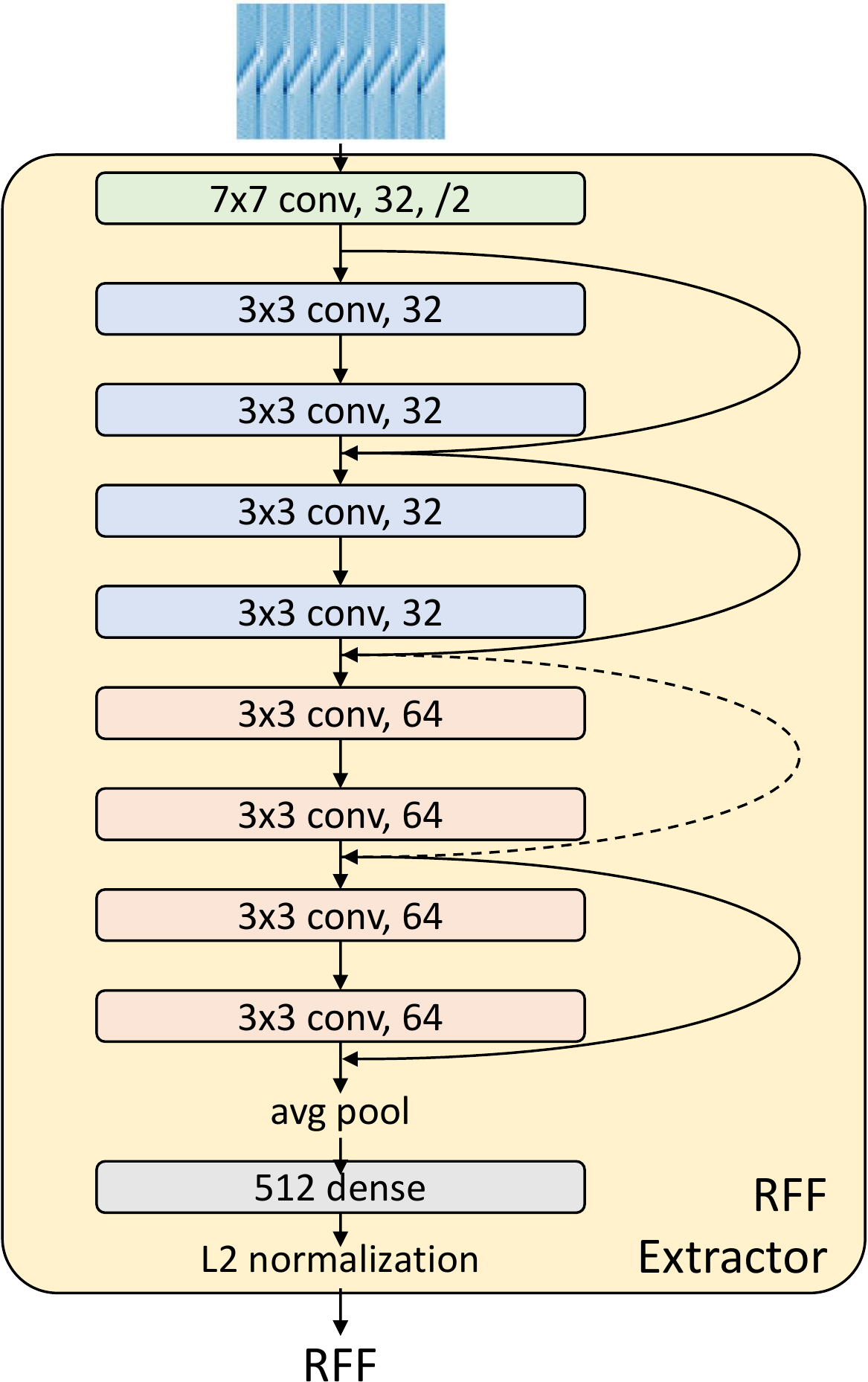}
	\caption{Architecture of the RFF extractor, revised from ResNet.}
	\label{fig:backbone}
\end{figure}

The architecture of the RFF extractor is shown in Fig.~\ref{fig:backbone}, where $/2$ denotes strides two. 
It consists of nine convolutional layers, one average pooling layer and one dense layer of 512 neurons, residual structure is also adopted. The first convolution layer uses 32 7$\times$7 filters with stride 2, the second to the fifth layers use 32 3$\times$3 filters, and the sixth to the ninth layers employ 64 3$\times$3 filters. All the convolutional layers are activated by rectified linear unit (ReLU) and padding is used. The L2 normalized model output is a vector consists of 512 elements which can be considered as the RFF extracted from the received packet.

%

%
\subsection{Deep Metric Learning}
%

\begin{figure}[!t]
	\centering
	\includegraphics[width=2.4in]{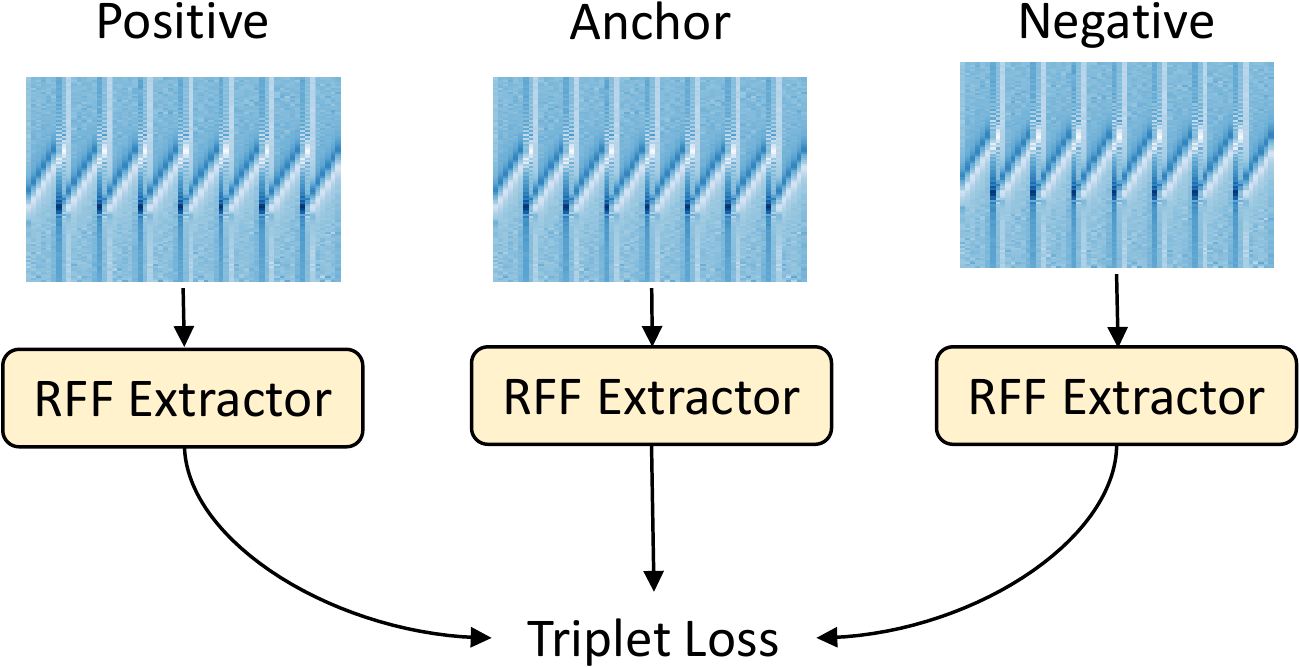}
	\caption{Triplet loss.}
	\label{fig:triplet_loss}
\end{figure}

Deep metric learning aims to project the input data into a space where similar samples are close to each other and dissimilar samples are far away.
Triplet loss is a well-known embedding loss used in deep metric learning and has been successfully adopted in face recognition~\cite{schroff2015facenet,musgrave2020metric}. During the training process, a triplet consisting of anchor, positive and negative samples is selected from the training set at each step. Specific to RFFI, anchor and positive samples are packets from the same device and the negative sample is from a different one. As shown in Fig.~\ref{fig:triplet_loss}, their RFFs are extracted so that the triplet loss can be calculated. 
The goal of triplet loss is to minimize the Euclidean distance between the anchor and positive sample while maximizing the distance between the anchor and negative sample, which can be written as
\begin{equation}
Loss = \max(D(Anc,Pos) - D(Anc,Neg) + \alpha, 0)
\end{equation}
where $Anc$, $Pos$, $Neg$ denote features extracted from anchor, positive and negative samples, respectively. $D(\cdot,\cdot)$ denotes the Euclidean distance between two vectors. $\alpha$ is a predefined hyperparameter which denotes the margin between positive and negative pairs. It is set to 0.1 in our implementation.


\subsection{Training Details}\label{sec:training_details}

The RFF extractors in this paper are all trained with the same settings, including validation set ratio, optimizer, learning rate schedule, batch size and stop condition. 10\% of the training data are randomly separated for validation. The model is optimized using RMSprop with an initial learning rate of 0.001. The learning rate drops every time the validation loss does not decrease for 10 epochs and the drop factor is 0.2. The batch size is set to 32. The training stops when validation loss does not decrease for 30 epochs. 
The model is implemented using Keras and trained with NVIDIA GeForce GTX 1660.

\section{Enrollment and Identification}\label{sec:enrollment_identification}

%

\subsection{Enrollment}
The legitimate devices are required to send several packets for enrollment before joining the system. These enrollment packets are first preprocessed and transformed to channel independent spectrograms. Then the RFF extractor is used to extract the RFF templates and store them in a database. 
The enrollment can be regarded as the training phase of a k-NN classifier, which simply memorizes all the training samples. 
We collect 100 packets from each device for enrollment, which is sufficient since k-NN is not a data-hungry model like deep learning. This is very desirable as it can significantly decrease the overhead of data collection.



\subsection{Identification}\label{sec:identification}

%
A complete identification system should include two parts, namely rogue device detection and device classification.
In our implementation, both of them are based on the k-NN algorithm with simple distance measures. The RFF extractor is trained with triplet loss which is defined by the Euclidean distance. The k-NN algorithm also relies on Euclidean distance measures thus their principles match well. 

\subsubsection{Rogue Device Detection}
Rogue device detection is to determine whether the received packet is from an enrolled legitimate device, which is an anomaly detection problem. It is necessary before device classification, otherwise, the RFFI system will assign legitimate labels even to rogue devices.

The rogue device detection is implemented by the distance-based k-NN anomaly detection algorithm. The RFF of the received packet is extracted and the average distance to its $K$ nearest neighbors in the RFF database is calculated as the detection score, which can be mathematically expressed as
\begin{equation}
D_{avg} = \frac{1}{K} \sum_{i=1}^{K} D_i ,
\end{equation}
where $D_i$ is the Euclidean distance to the i-th neighbor. Then a predefined threshold $\lambda$ is used to determine whether the received packet is from an enrolled device. When the detection score $D_{avg}$ is above $\lambda$, the packet is considered to be sent from a rogue device and the authentication is failed. In contrast, the packet is considered to come from an enrolled device when $D_{avg}$ is below the threshold $\lambda$ and will be further classified. This can be formulated as
\begin{equation}
Decision = \left\{
\begin{aligned}
\mbox{enrolled device}, when \; D_{avg} \leq  \lambda \\
\mbox{rogue device}, when \; D_{avg} > \lambda
\end{aligned}
\right.
\end{equation}

The evaluation metric of rogue device detection is the receiver operating characteristic (ROC) curve. It reveals the trade-off between false-positive rate (FPR) and true-positive rate (TPR) at various threshold settings. The area under the curve (AUC) calculated from the ROC curve is another important evaluation metric. The closer it is to one, the better the detection performance is.

%

\subsubsection{Device Classification} 
Device classification is to infer the specific label of the transmitter from which the received packet is sent, which is a classification problem. It outputs a previously enrolled label according to the templates stored in the RFF database. 
 
The proposed device classification system is implemented with the majority voting k-NN algorithm. The RFF of the received packet is first extracted and its $K$ nearest neighbors are selected from the database according to the Euclidean distance. This packet is then assigned to the label that is most frequent among the $K$ neighbors. 

The evaluation metrics for a device classification system are the confusion matrix and overall accuracy that is defined as the correctly classified samples divided by the total number of test samples. The number of neighbors $K$ is set to 15 for both rogue device detection and device classification.

\section{Experimental Evaluation}\label{sec:experimental_evaluation}

\subsection{Experimental Settings}

\begin{figure}[!t]
	\centering
	\subfloat[]{\includegraphics[width=1.1in]{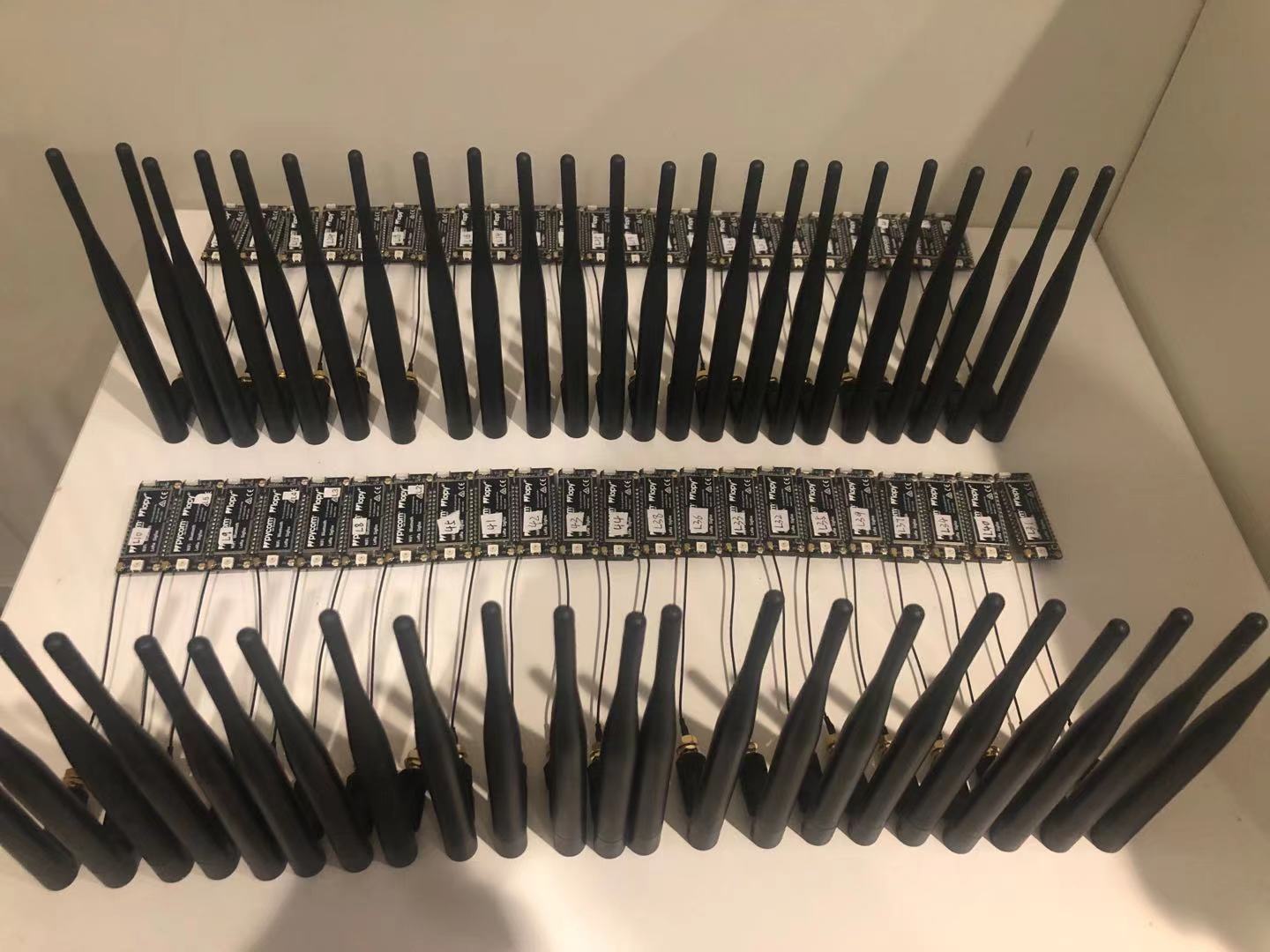}
		\label{fig:dut1_45_photo}}
	\subfloat[]{\includegraphics[width=1.1in]{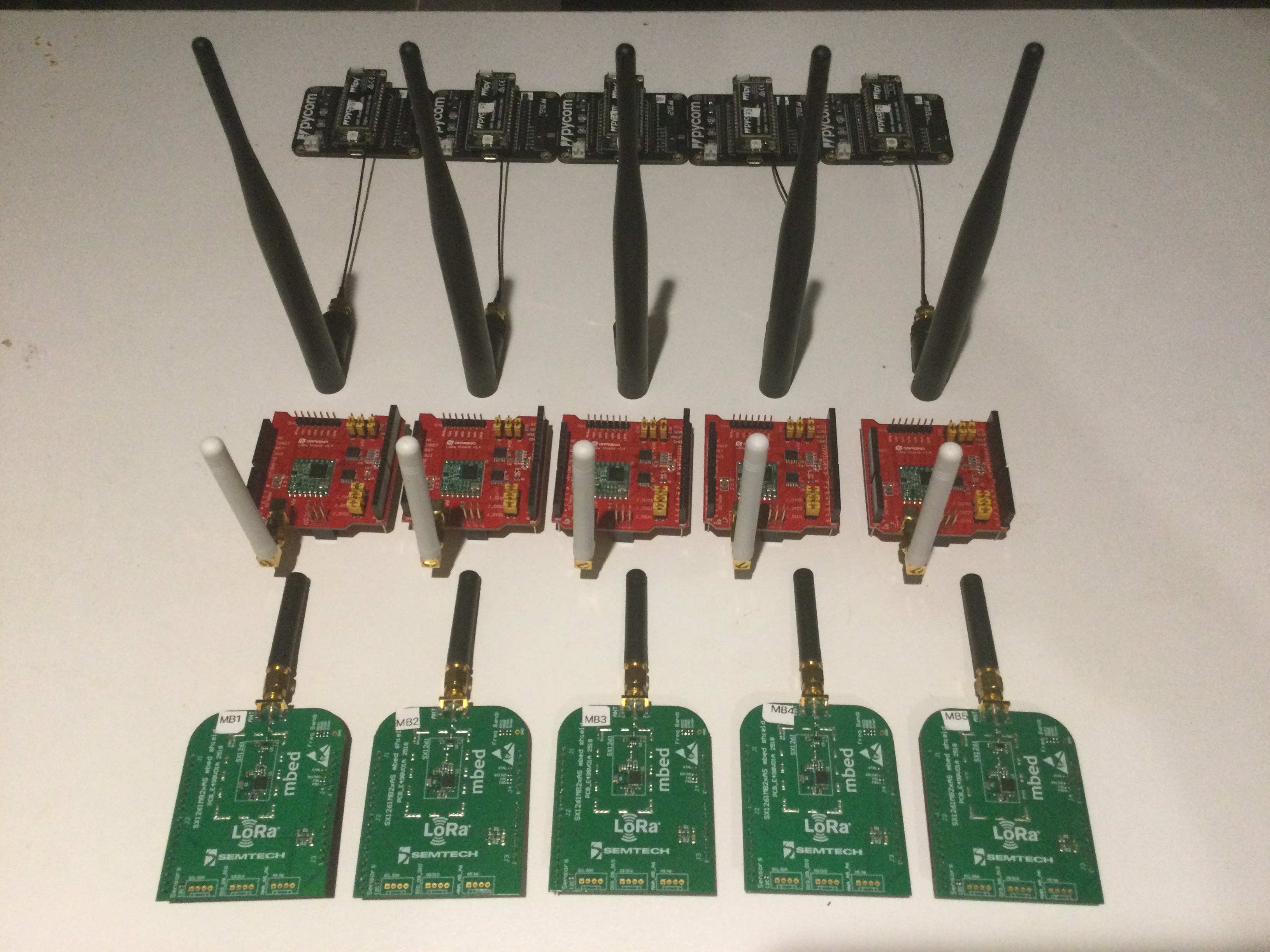}
		\label{fig:dut46_60_photo}}	
	\subfloat[]{\includegraphics[width=1.1in]{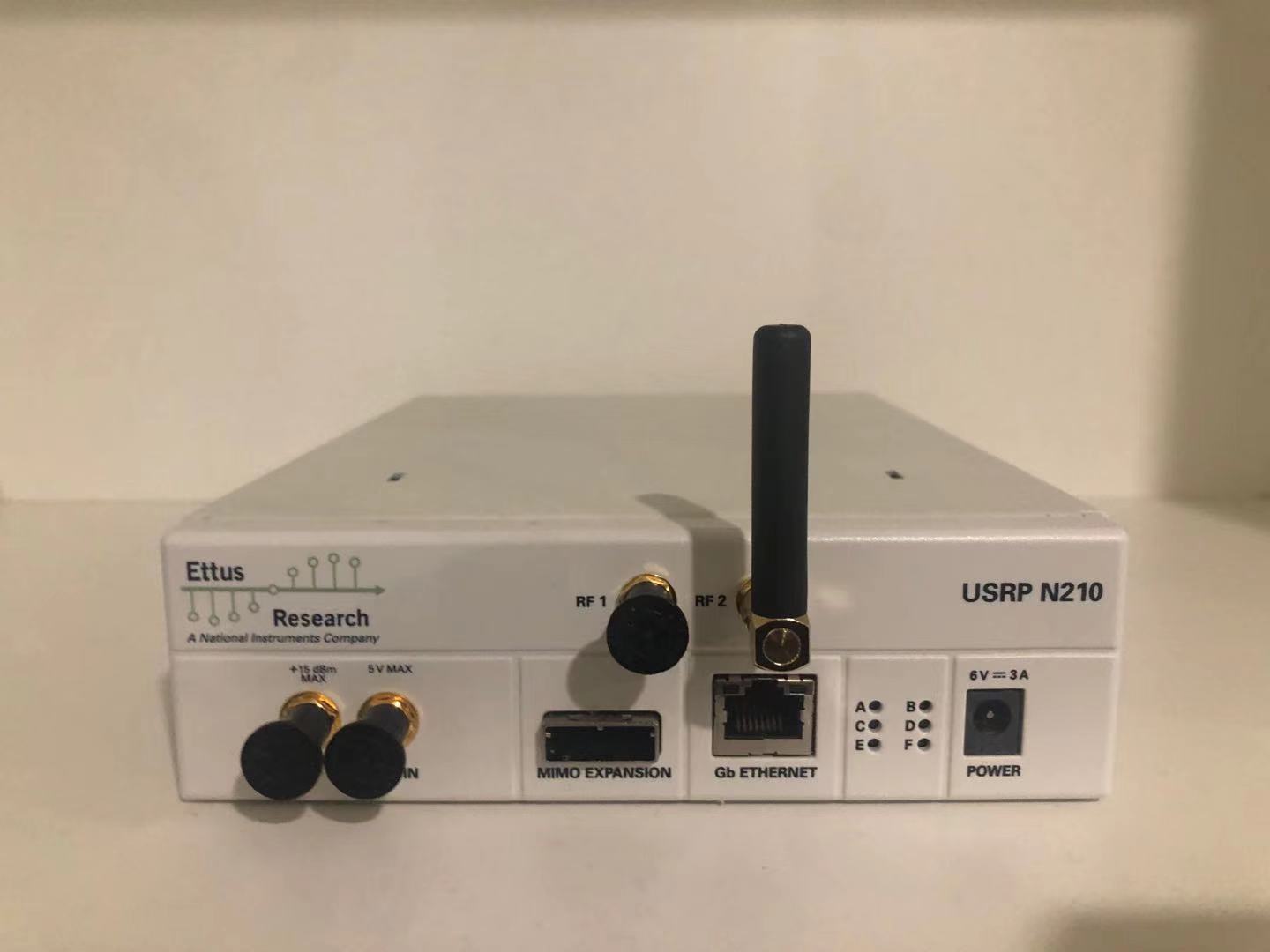}
		\label{fig:usrp_photo}}
	\caption{Experimental devices. (a) DUTs 1-45: LoPy4. (b) DUTs 46-60: mbed SX1261 shields, FiPy and Dragino SX1276 shields. (c) USRP N210 SDR.}
	\label{fig:experimental_equipment}
\end{figure}

We employed 60 commercial off-the-shelf LoRa devices as devices under test (DUTs), and a USRP N210 software defined radio (SDR) platform as the receiver, as shown in Fig.~\ref{fig:experimental_equipment}. 
Detailed DUTs information can be found in Table~\ref{tab:dut_information}. 
\begin{table}[!t]
	\centering
	\caption{LoRa DUTs.}
	\begin{tabular}{|l|l|l|}
		\hline
		DUT index & Model  & Chipset \bigstrut\\
		\hline
		1 - 45   & Pycom LoPy4 & SX1276 \bigstrut\\
		\hline
		46 - 50 & mbed SX1261 shield & SX1261 \bigstrut\\
		\hline
		51 - 55 & Pycom FiPy & SX1272 \bigstrut\\
		\hline
		56 - 60 & Dragino SX1276 shield & SX1276 \bigstrut\\
		\hline
	\end{tabular}%
	\label{tab:dut_information}%
\end{table}%
The carrier frequency is 868.1~MHz and the transmission interval is set to 0.3~s. 

Table~\ref{tab:summary_subsections} summarizes our experimental studies and their configurations.
\begin{itemize}
\item Training and RFF extractors: We first collect 500 packets from each of DUT 1-30 in a residential room with LOS between the transmitter and receiver.  The distance is about half a meter and DUT antennas are vertical to the ground. \textbf{Extractor 1} is trained as a baseline model involving channel independent spectrogram and data augmentation. \textbf{Extractors 2-6} are trained for comparison to show the effects of different procedures during training. The detailed extractor information can be found in Table~\ref{tab:extractor_information}. \textbf{Extractor 1-3} are trained with the dataset augmented by both multipath and Doppler effects, while during the training of \textbf{Extractor 6} only the multipath effect is emulated. 
\begin{table}[!t]
  \centering
  \caption{Extractor information.}
    \begin{tabular}{lp{6.145em}ll}
    \hline
      & Extractor \newline{}input & \multicolumn{1}{p{4.07em}}{Training \newline{}DUTs} & \multicolumn{1}{p{3.215em}}{Data\newline{}augmentation} \bigstrut\\
    \hline
    Extractor 1 & Channel ind\newline{}Spectrogram & DUT 1-30 & Yes \bigstrut\\
    \hline
    Extractor 2 & Channel ind.\newline{}Spectrogram & DUT 1-20 & Yes \bigstrut\\
    \hline
    Extractor 3 & Channel ind. \newline{}Spectrogram & DUT 1-10 & Yes \bigstrut\\
    \hline
    Extractor 4 & Channel ind.\newline{}Spectrogram & DUT 1-30 & No \bigstrut\\
    \hline
    Extractor 5 & \multicolumn{1}{l}{Spectrogram} & DUT 1-30 & No \bigstrut\\
    \hline
    Extractor 6 & Channel ind.\newline{}Spectrogram & DUT 1-30 & Yes, $f_d$=0~Hz \bigstrut\\
    \hline
    \end{tabular}%
  \label{tab:extractor_information}%
\end{table}%

\item Enrollment: We collect 100 packets from each DUT for enrollment. Unless otherwise specified, the enrollment sets are collected in a residential room with LOS and vertical antenna placement.
\item Identification: 100 packets are collected from each DUT. The experimental setup varies according to the tests.
\end{itemize}
The RFF extractors only need to be trained once. Enrollment and identification are carried out multiple times for evaluating system performance in various configurations.

\begin{table*}[!t]
  \centering
  \caption{Summary of Experimental Evaluation.}
    \begin{tabular}{|l|r|p{15em}|p{13em}|p{18em}|}
    \hline
    Section & \multicolumn{1}{l|}{Extractor} & \multicolumn{1}{l|}{Enrollment Set} & \multicolumn{1}{l|}{Identification Set} & \multicolumn{1}{l|}{Purpose} \bigstrut\\
    \hline
    Section 7.3 & 1,2,3 & Seen, unseen, and different model DUTs (residential room) & Seen, unseen, and different model DUTs (residential room) & Generalization ability of device classification versus the number of training DUTs \bigstrut\\
    \hline
    Section 7.4 & 1,2,3 & Legitimate DUTs (residential room) & Legitimate  and rogue DUTs \newline{}(residential room) & Generalization ability of rogue device detection versus the number of training DUTs \bigstrut\\
    \hline
    Section 7.5 & 1,4,5 & Stationary dataset (residential room) & Ten datasets with different channel conditions (office building) & Evaluate data augmentation and channel independent spectrogram  \bigstrut\\
    \hline
    Section 7.6 & 1,6 & Stationary dataset (residential room) & Emulated datasets with various moving speeds & Evaluate data augmentation (Doppler effect) \bigstrut\\
    \hline
    Section 7.7 & 1  & Datasets with different antenna directions (office) & Datasets with different antenna directions (office) & Effect of antenna polarization \bigstrut\\
    \hline
    \end{tabular}%
  \label{tab:summary_subsections}%
\end{table*}%

\subsection{System Scalability}\label{sec:system_scalability}

RFFI system should be scalable to allow devices to join and leave. Previous deep learning-based RFFI systems use a single classification neural network therefore the model output can only be device labels that are present during training. For example, if we train with the packets collected from DUTs 1-10, the model can only output label 1-10. Therefore, the deep learning model must be retrained when a new DUT 11 joins the system, which is time-consuming and not practical.

In this paper, we train the deep learning model as an RFF extractor rather than for classification. The training only needs to be done once. We then introduce an enrollment stage to obtain RFFs of any devices in the IoT network via the pre-trained RFF extractor. The enrollment is training-free and can be done very quickly when new devices join. The RFF database can also be managed efficiently by only recording the RFF of devices that are active and present in the IoT network. In summary, our protocol has excellent scalability in terms of maintaining an up-to-date device list.

\subsection{Generalization Ability for Device Classification}\label{sec:generalization_evaluate}
To achieve good scalability, the RFF extractor must be able to extract RFFs from the newly added devices that are out-of-library during the training stage. In other words, the RFF extractor should have an excellent generalization ability on previously unseen devices. The rule of thumb in deep metric learning is that the more training data, the better the generalization ability. We select \textbf{Extractor 1, 2, 3} for comparison since they are trained with different numbers of DUTs.


We specially select three groups of DUTs for evaluation, namely DUTs 1-10, DUTs 31-40 and DUTs 46-60. 
\begin{itemize}
	\item DUTs 1-10 are present during the training of \textbf{Extractor 1, 2, 3} therefore they are used to evaluate the extractor performance on seen devices. 
	\item DUTs 31-40 are disjoint with the training devices  but with the same model (LoPy4). They can be used to validate the extractor performance on unseen devices. 
	\item DUTs 45-60 are produced by other manufacturers whose hardware characteristics are much more different from the training devices. These DUTs require a higher generalization ability of the RFF extractor.
\end{itemize}

\begin{figure}[!t]
	\centering
	\includegraphics[width=3.2in]{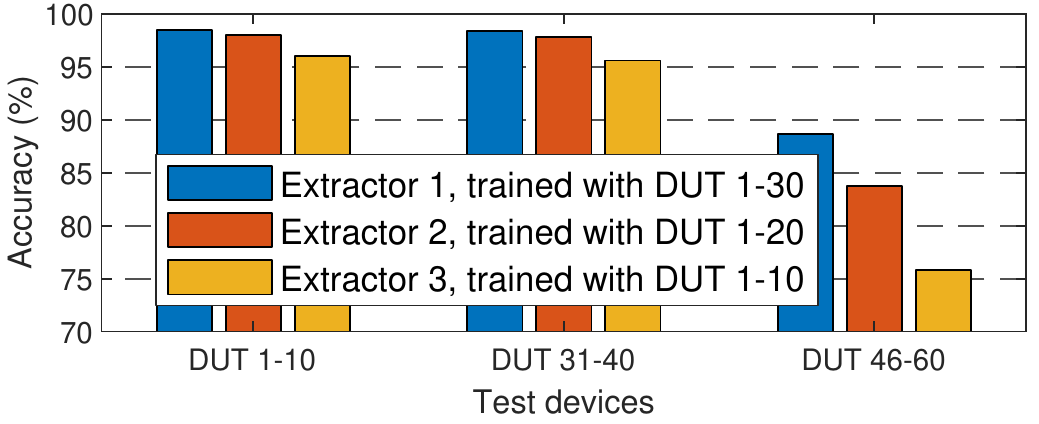}
	\caption{Performance of RFF extractors.}
	\label{fig:generalization_evaluate}
\end{figure}

The classification results on these three groups are shown in Fig.~\ref{fig:generalization_evaluate}.
It can be observed that \textbf{Extractor 1, 2, 3} perform excellently on DUTs 1-10 and DUTs 31-40. The overall accuracies are always above 95\%.
The highest accuracy is reached by \textbf{Extractor 1} with 98.50\% on DUTs 1-10. 
The accuracy is 98.40\% on classifying DUTs 31-40 which demonstrates the trained RFF extractor can efficiently extract RFFs from the devices that are not present during training.

\begin{figure}[!t]
	\centering
	\includegraphics[width=3.3in]{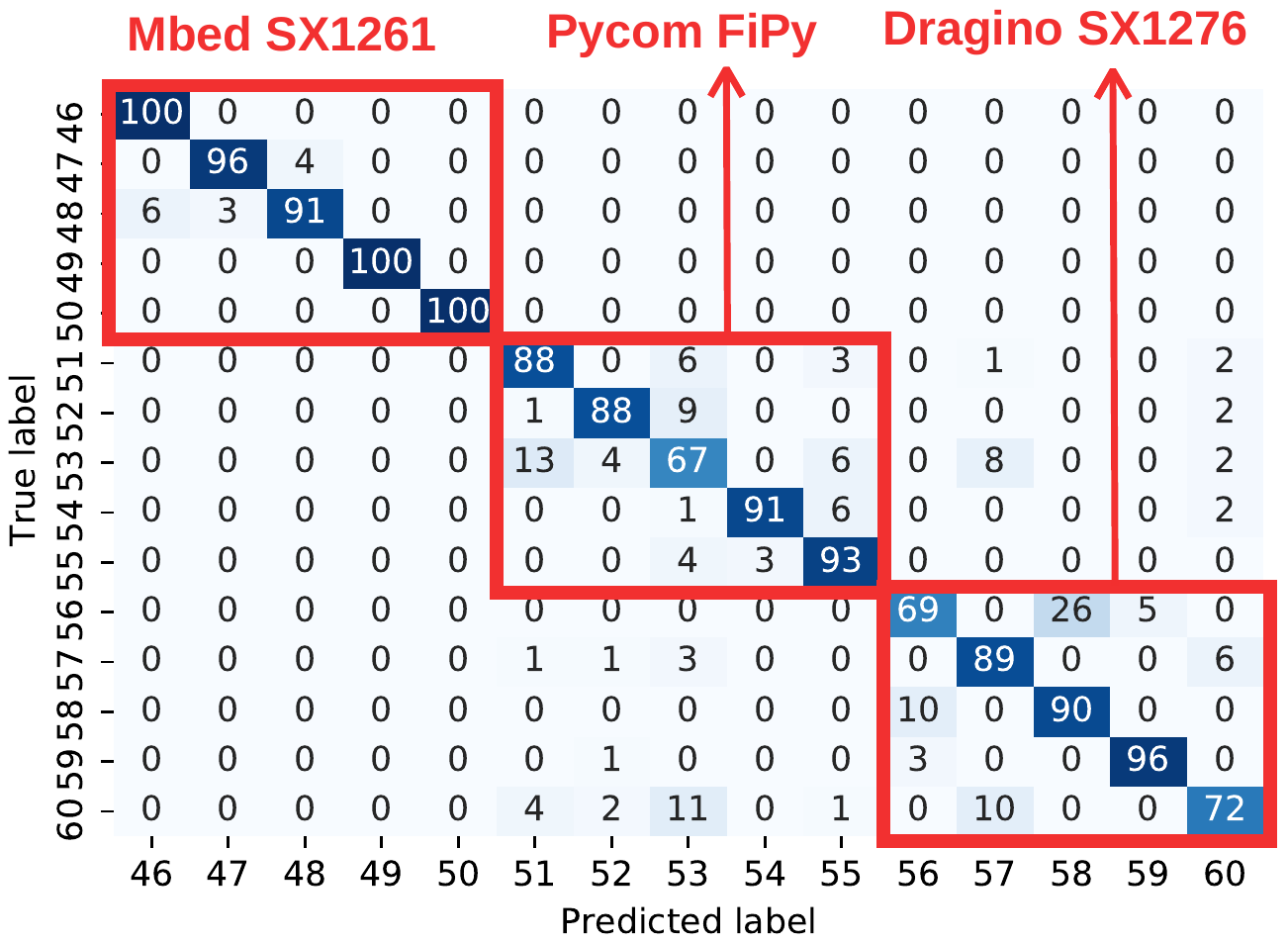}
	\caption{Classification result on other LoRa types, overall accuracy 88.67\%. }
	\label{fig:result_other_device_type}
\end{figure}

We can see that there is a significant performance gap between \textbf{Extractor 1, 2, 3} on DUT 46-60. \textbf{Extractor 3},  trained with only 10 DUTs, has the worst classification result, i.e., 75.80\%. Training with 30 devices (\textbf{Extractor 1}) can increase it to
88.67\% and the confusion matrix is shown in Fig.~\ref{fig:result_other_device_type}.
It is found that the DUTs of the same type have similar hardware characteristics as almost all the misclassified packets fall into the three red boxes. The classification performance on DUT 46-50 is quite excellent but that on DUT 56-60 is not satisfying. This may because the hardware characteristics of DUT 56-60 (Dragino SX1276) are more different from the training DUTs (LoPy4). 
In summary, more devices should be included for training to achieve good generalization ability on out-of-library devices. 


\subsection{Generalization Ability for Rogue Device Detection}\label{sec:evaluation_rogue_device_detection}

Rogue devices are also out-of-library devices that are inaccessible during training. Whether the number of training DUTs affects the performance of rogue device detection should be evaluated. The hardest case for rogue device detection is the rogue device has similar hardware characteristics to the legitimate ones. Therefore, we specially select DUT 31-40 as legitimate devices and DUT 41-45 as rogue ones, which are are all LoPy4 devices. Then we collect 100 packets from each of DUT 31-45 for evaluation.

\begin{figure}[!t]
	\centering
	\includegraphics[width=3in]{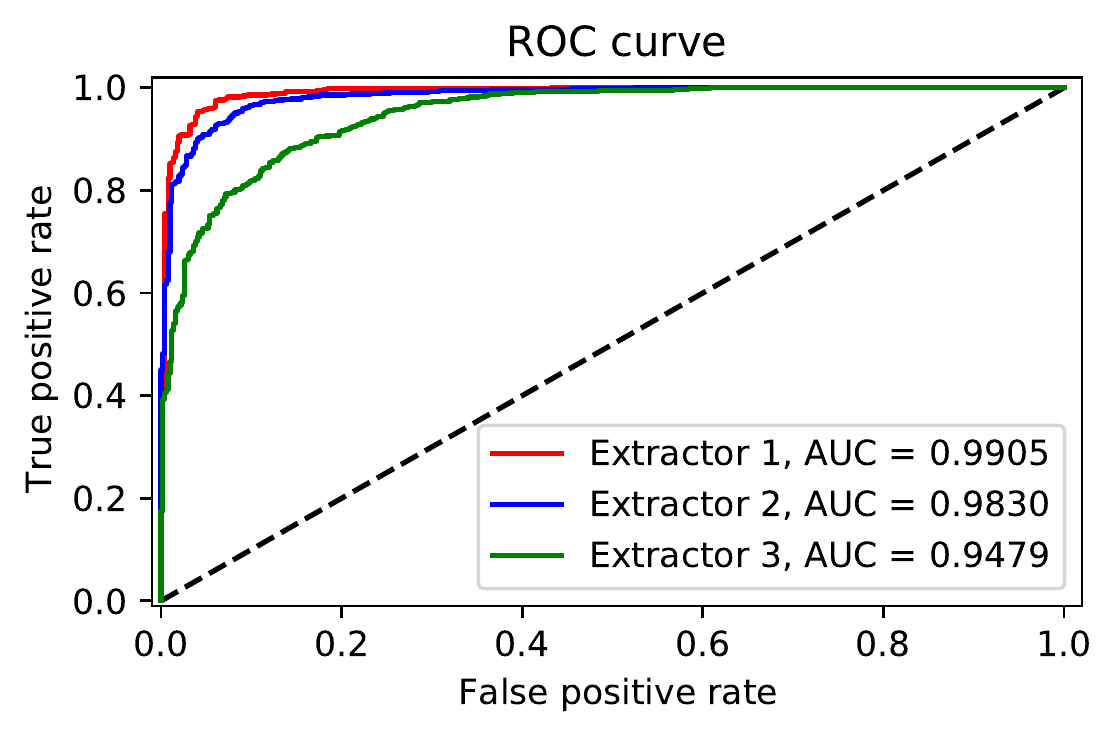}
	\caption{ROC curve of rogue device detection.}
	\label{fig:roc_curve}
\end{figure}

As discussed in Section~\ref{sec:identification}, the evaluation metric of rogue device detection is the ROC curve and corresponding AUC. 
The results are shown in Fig.~\ref{fig:roc_curve}.
It can be observed that the AUC of \textbf{Extractor 1} is 0.9905, indicating excellent detection performance. The AUC of \textbf{Extractor 3} is the worst, which is 0.9479. This demonstrates the more DUTs involved in the training stage, the better the rogue device detection is.

\subsection{Effect of Data Augmentation and Channel Independent Spectrogram}\label{sec:effect_wireless_channel}

\begin{figure}[!t]
	\centering
	\subfloat[]{\includegraphics[width=1.4in]{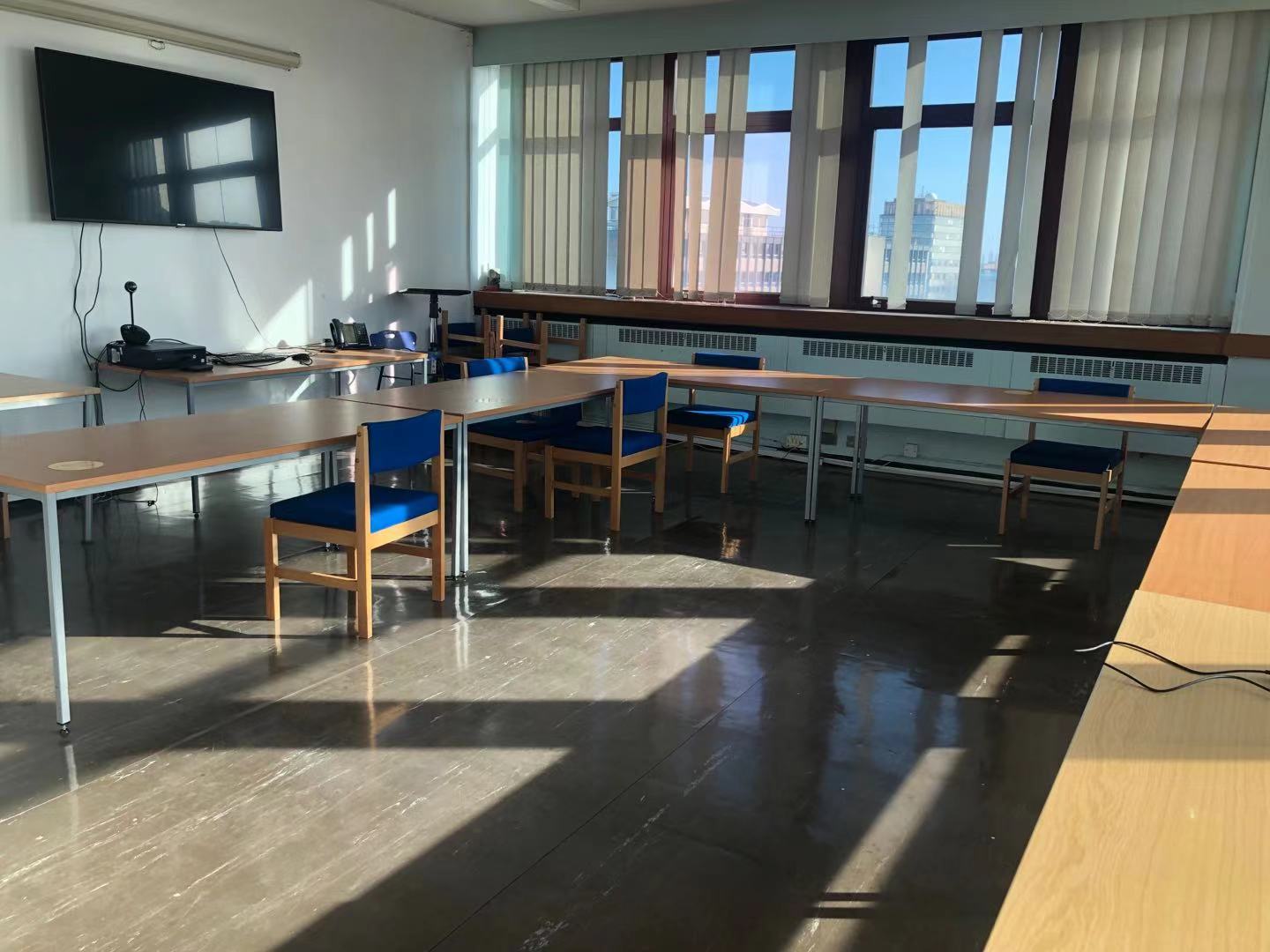}
		\label{}}
	\subfloat[]{\includegraphics[width=1.4in]{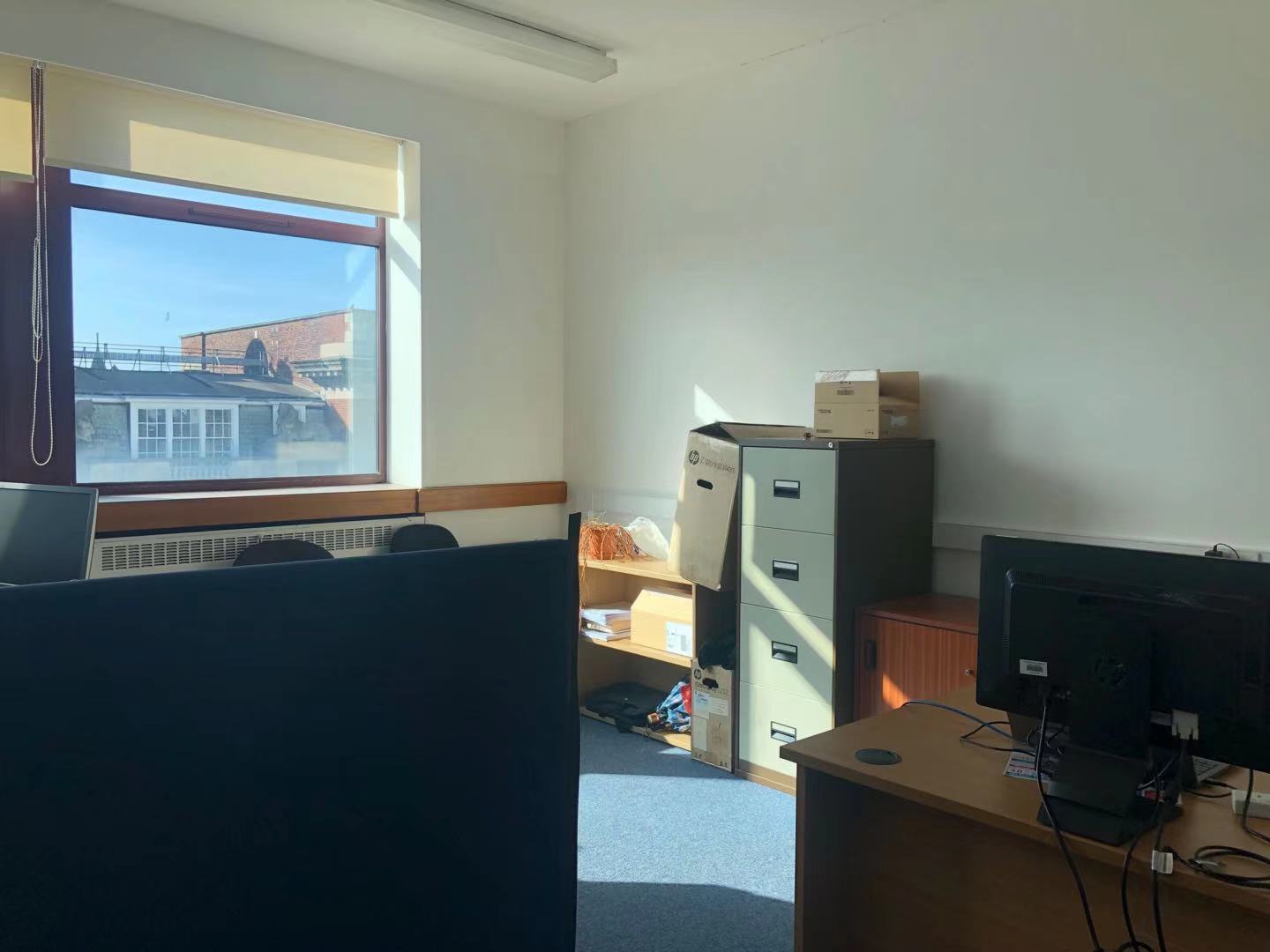}
		\label{}}
	
	\subfloat[]{\includegraphics[width=3.4in]{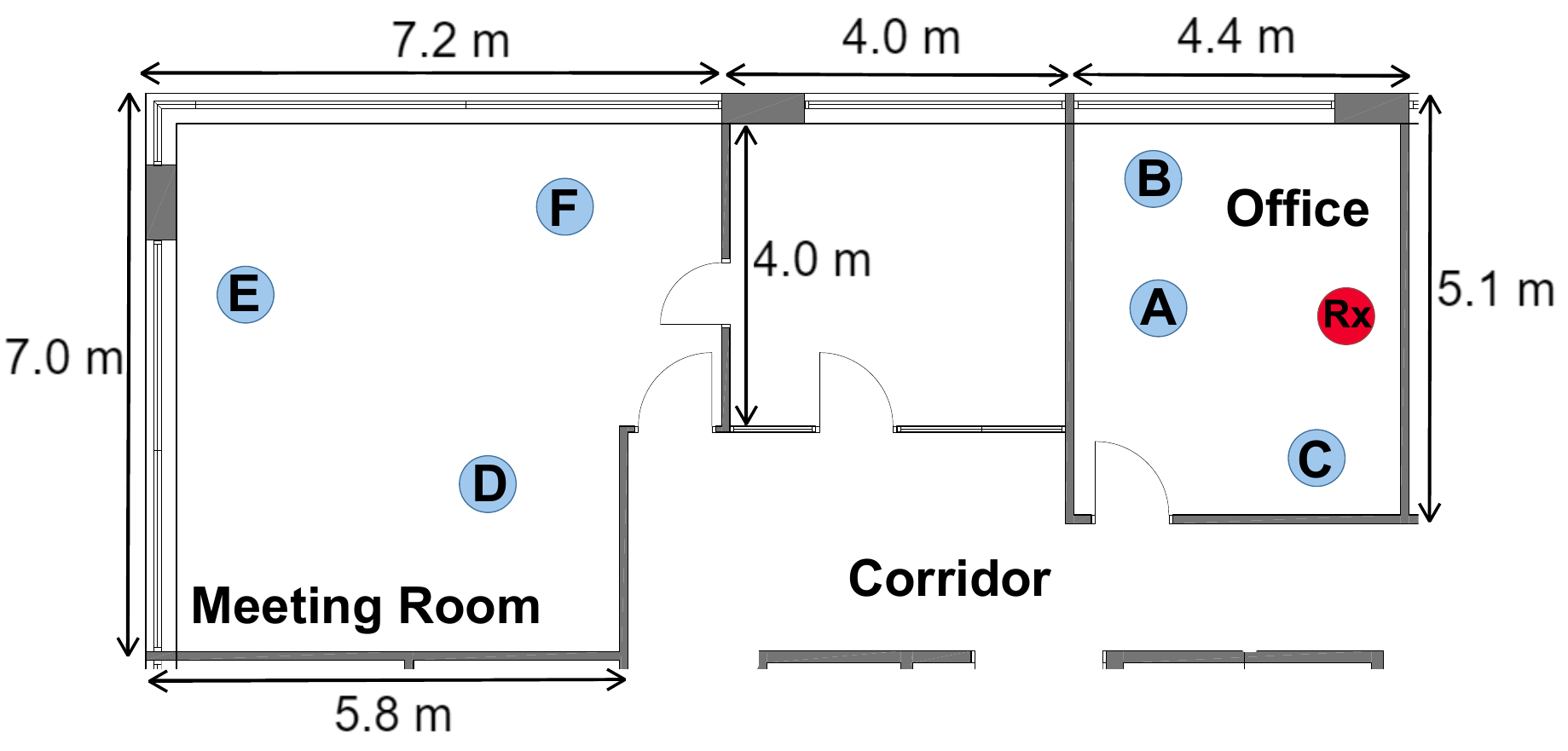}}	
	\caption{Experimental environments. (a) Meeting room. (b) Office. (c) Floor plan.}
	\label{fig:environment}
\end{figure}

The RFFI system should be robust to locations and channel variations. We mitigate the channel effect in the time-frequency domain and propose the channel independent spectrogram as the model input. Then we use data augmentation to further increase its robustness to the wireless channel. To verify that they are effective, we collect identification datasets in an office building whose environment is completely different from the enrollment residential room. The experiments are conducted in an office and a meeting room. The photos and floor plan of them can be found in Fig.~\ref{fig:environment}.
We collect 10 identification datasets, \textbf{D1-10}, in total and they can be categorized into three scenarios, namely stationary, object moving and mobile scenarios.

\subsubsection{Stationary Scenario}
Six datasets, \textbf{D1-6}, are collected at Location A-F. During the data collection, DUTs and USRP N210 are stationary and there is no object moving around. 

Locations A-F lead to multipath effects of varying severity. For instance, the waveform of DUT 6 at Location~F (\textbf{D6}) is similar to that shown in Fig.~\ref{fig:DUT6_LocF}, showing a distinct sawtooth shape. While the waveform at Location A is almost the same as Fig.~\ref{fig:DUT6_LocA}, which is nearly flat. As discussed in Section~\ref{sec:lora_channel}, the sawtooth is caused by the multipath effect therefore the channels at Locations A and F are different.

\subsubsection{Object Moving Scenario}
Two datasets, \textbf{D7~\&~D8}, are collected at location B and F, respectively. The DUTs and USRP are kept stationary while a person randomly walks around the office at a speed of 2~m/s. 

In this scenario, the collected LoRa packets show different waveforms since the channel changes as a result of people walking. However, the Doppler effect is not quite serious due to the low walking speed.

\subsubsection{Mobile Scenario}
Two datasets, \textbf{D9~\&~D10}, are collected  in the office and meeting room, respectively. The USRP is kept stationary and a person takes the DUTs walking around at a speed of 2~m/s. 

Both multipath and Doppler effects are serious in this scenario since the sawtooth shapes and amplitude variations can be frequently observed. Many packets show waveforms similar to those shown in Fig.~\ref{fig:DUT1_walking_office} and Fig.~\ref{fig:DUT1_walking_meeting_room}. 


\subsubsection{Discussion}
\begin{figure*}[!t]
	\centering
	\includegraphics[width=7in]{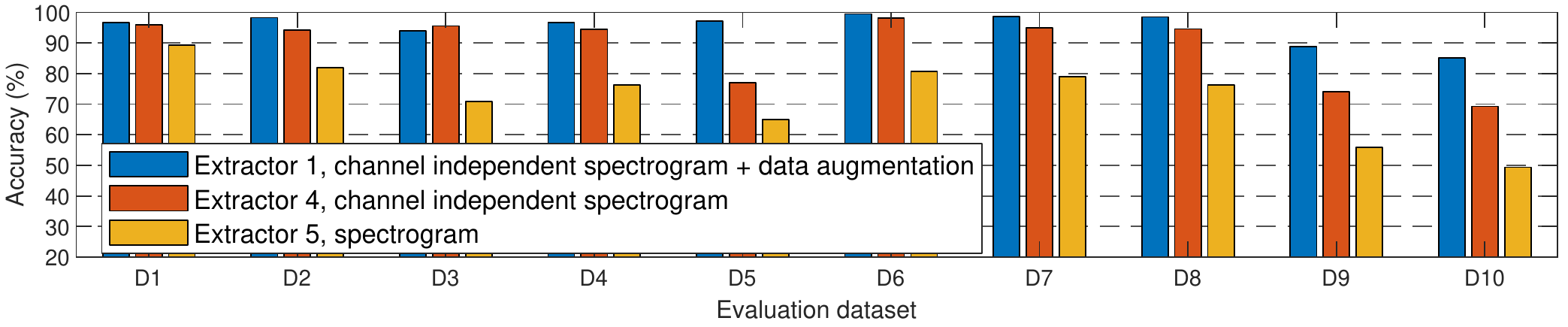}
	\caption{Classification results in various channel conditions.}
	\label{fig:channel_result}
\end{figure*}

\textbf{Extractor 1, 4, 5} are selected for comparison to demonstrate the effectiveness of channel independent spectrogram and data augmentation. Their training details can be found in Table~\ref{tab:extractor_information} and the classification results are summarized in Fig.~\ref{fig:channel_result}.

It can be observed that \textbf{Extractor 1} performs well on all datasets, which indicates the proposed RFFI system is robust to locations and channel variations. 
It performs slightly worse on \textbf{D9} and \textbf{D10} that are collected in mobile scenarios (lower than 90\%). This is possibly due to the inevitable shaking of the DUT antenna during moving, which will result in the change of antenna polarization. The effect of antenna polarization is discussed in Section~\ref{sec:antenna_polarization}.

Compared with \textbf{Extractor 1}, the training of \textbf{Extractor~4} does not employ data augmentation. As can be seen, its performance on \textbf{D6} is significantly worse with only 78.80\% accuracy. \textbf{D6} is collected at Location E which is the farthest from the receiver. This indicates the channel independent spectrogram is affected by the noise in lower SNR scenarios. Data augmentation must be used to further improve its robustness.

As for \textbf{Extractor 5}, it employs neither  channel independent spectrogram nor the data augmentation. It can be observed that \textbf{Extractor 5} only performs well on \textbf{D1} which is a short-distance LOS stationary scenario. It is presumed that the channel condition at Location A is similar to the enrollment residential room. Once the channel is changed, the classification performance will degrade significantly. The worst results occur on \textbf{D9} and \textbf{D10} (mobile scenario) whose accuracies are 55.90\% and 49.40\%, respectively, which are 30\% lower than \textbf{Extractor 1}.


We also try to explore whether data augmentation can solve the channel problem without the employment of channel independent feature. However, we found out the deep learning model cannot converge as the training loss does not decrease at all. In other words, the RFF extractor is not capable to extract channel independent RFFs from the spectrogram when training data is distorted by  wireless channels. 
In contrast, when we use channel independent spectrogram instead of the spectrogram as model input (\textbf{Extractor 1}), the channel effects are mitigated and device-specific features can be learned so that the model converges.

\subsection{Effect of Doppler Shift in Data Augmentation}\label{sec:doppler_data_augmentation}

Doppler effect must be considered during data augmentation, especially in high-speed scenarios.
Due to the experimental constraints, we emulate the communication scenarios at various moving speeds by filtering the real collected \textbf{D1} to a channel simulator to generate identification sets. The process is almost the same with data augmentation introduced in Section~\ref{sec:data_augmentation}. The maximum Doppler frequency, $f_d$, is fixed to 0, 10, 30, 50, 100~Hz, respectively. These Doppler frequencies are equivalent to moving speeds of 0, 12.46, 37.33, 62.21 and 124.4~km/h for a LoRa system operating at 868 MHz. 
We select $\textbf{Extractor 1}$ and $\textbf{Extractor 6}$ for comparison to demonstrate the Doppler effect must be considered during data augmentation. 

\begin{figure}[!t]
	\centering
	\includegraphics[width=3.2in]{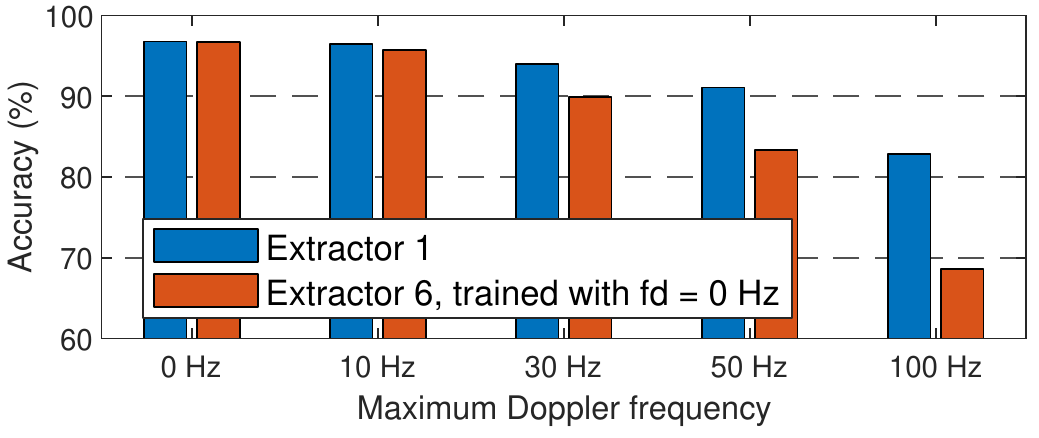}
	\caption{Classification results under various Doppler frequencies of the identification sets.}
	\label{fig:doppler_result}
\end{figure}

The results are summarized in Fig.~\ref{fig:doppler_result}.
Compared with $\textbf{Extractor 1}$, $\textbf{Extractor 6}$ does not consider the Doppler effect during data augmentation. Its maximum Doppler frequency $f_d$ is fixed to 0~Hz. It can be seen that these two extractors perform nearly the same in stationary scenarios ($f_d$ = 0~Hz). However, as the $f_d$ increases, their performance gap becomes wider. In the scenario of $f_d$ = 100 Hz, the performance of \textbf{Extractor 6} is only 68.60\% while \textbf{Extractor 1} is still above 80\%. The performance gap between them indicates the Doppler effect must be included in data augmentation to support the excellent performance in high-speed scenarios. 




\subsection{Effect of Antenna Polarization}\label{sec:antenna_polarization}
Antenna polarization refers to the direction of the electric field produced by an antenna. It was found to affect the transient-based RFFI system~\cite{danev2009transient}. To the best knowledge of the authors, there is no study on antenna polarization in deep learning-based approaches that use the entire signal for identification. To explore this, we additionally collect two datasets, \textbf{D11~\&~D12}, at Location~B and Location~F, respectively. The DUT antenna is parallel to the ground and points towards the USRP. Both DUTs and USRP are kept stationary.  \textbf{Extractor 1} is used to extract RFFs.

\begin{figure}[!t]
	\centering
	\subfloat[]{\includegraphics[width=1.6in]{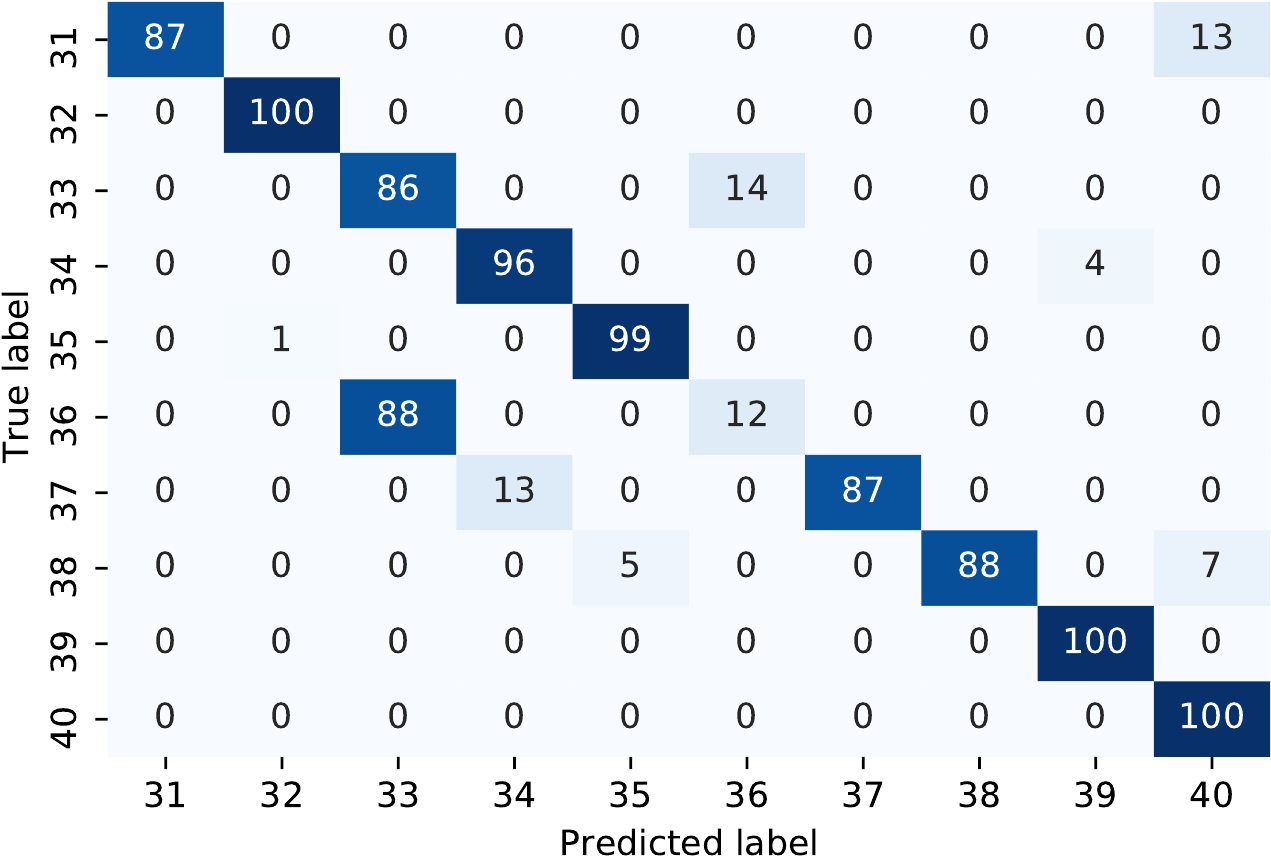}
		\label{fig:B_antenna}}
	\subfloat[]{\includegraphics[width=1.6in]{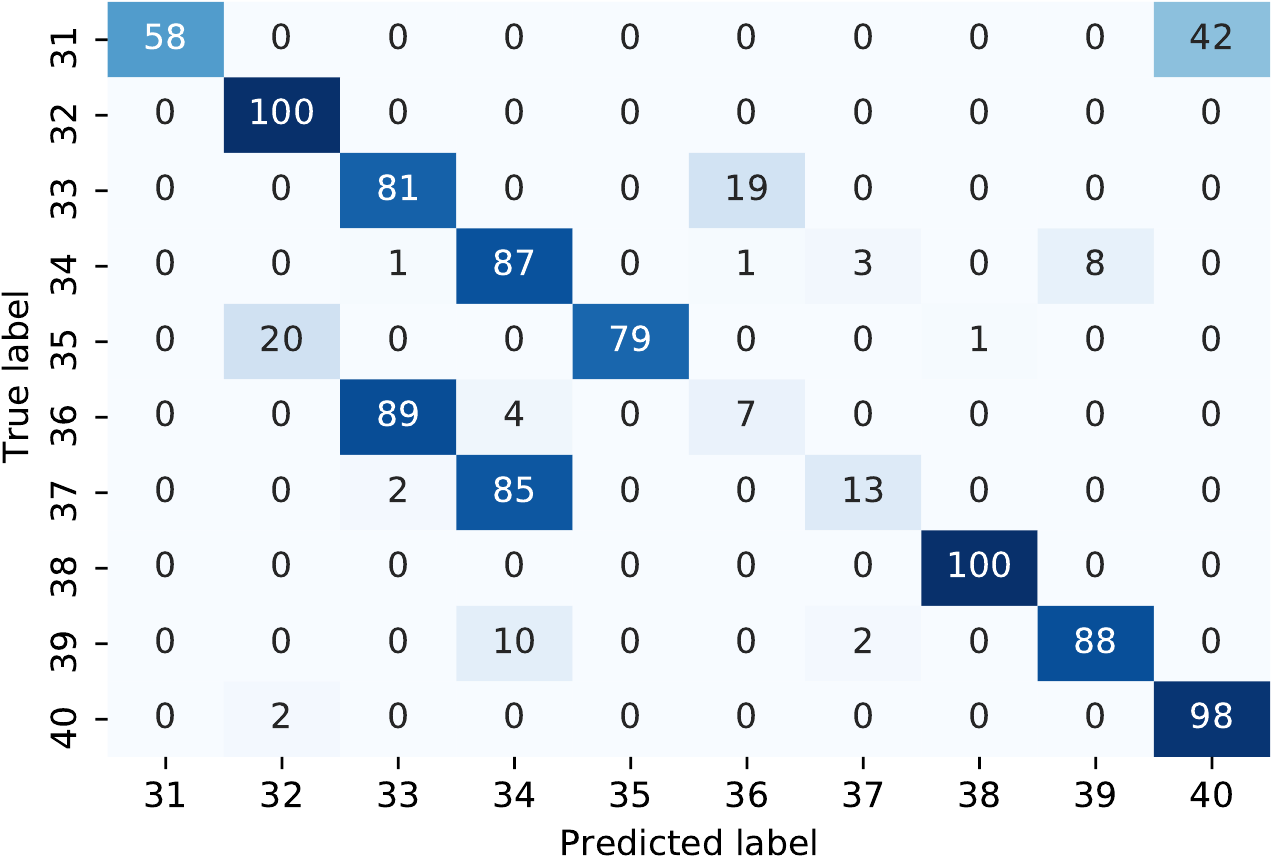}
		\label{fig:G_antenna}}
	
	\subfloat[]{\includegraphics[width=1.6in]{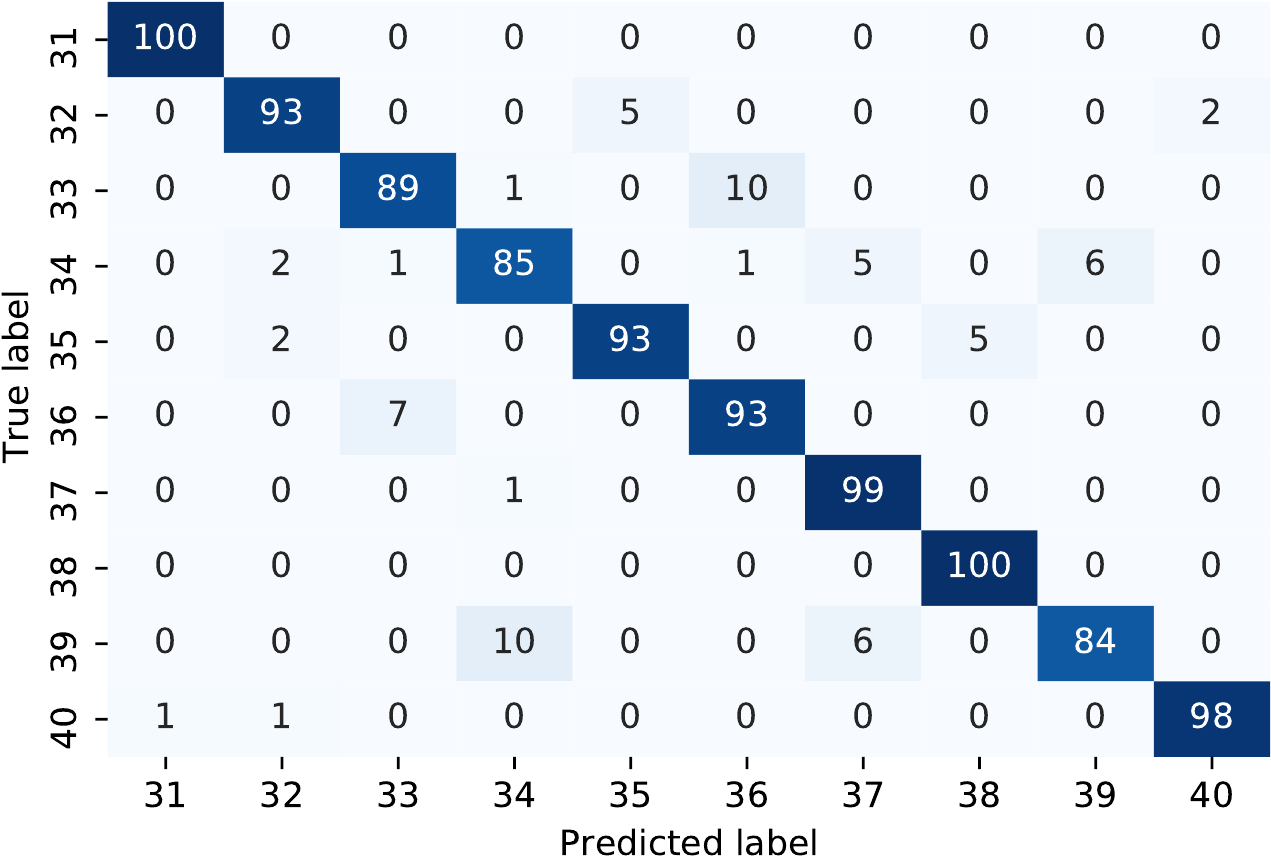}
		\label{fig:B_G_antenna}}
	\subfloat[]{\includegraphics[width=1.6in]{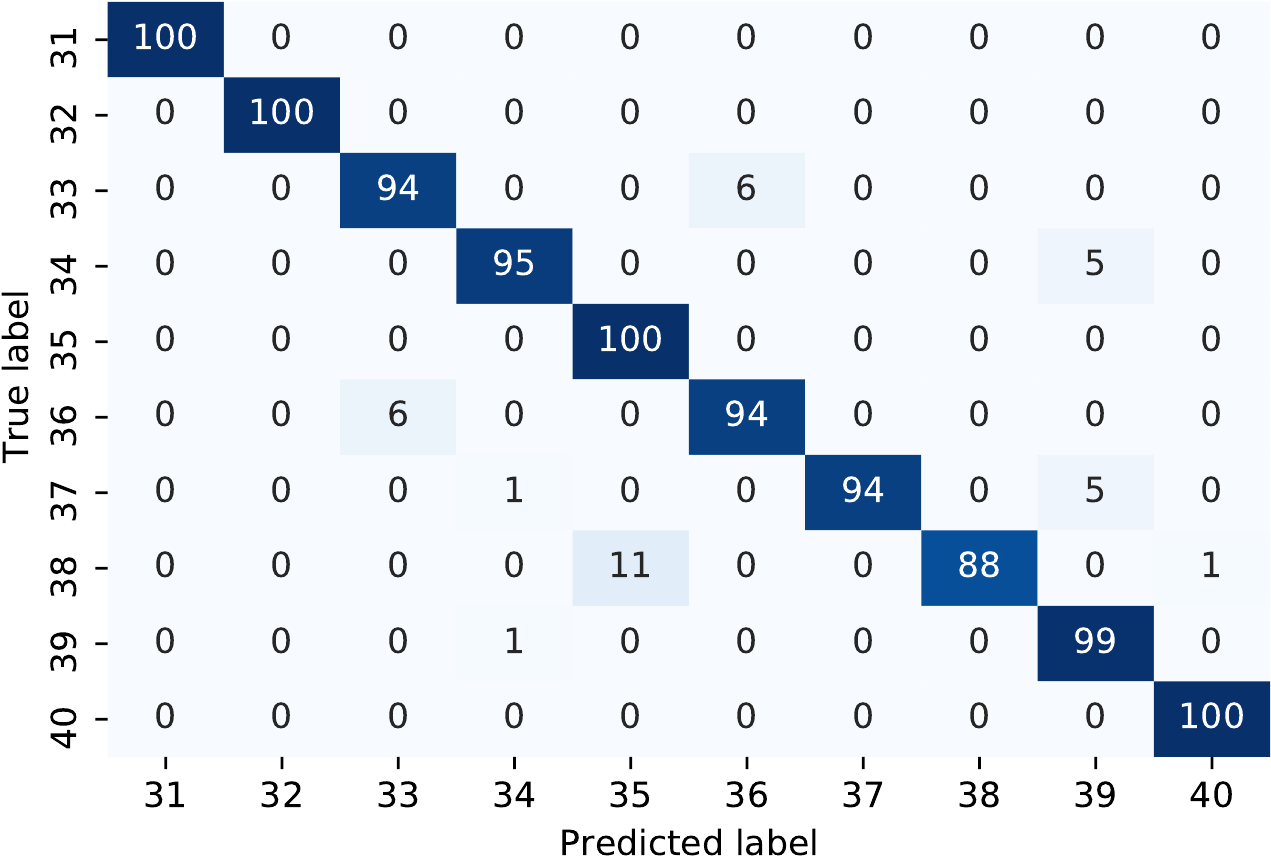}
		\label{fig:G_B_antenna}}
	\caption{Classification results of antenna polarization. (a) Enrollment \textbf{D2}, evaluation \textbf{D11}, overall accuracy 85.50\%. (b) Enrollment \textbf{D6}, evaluation \textbf{D12}, overall accuracy 71.10\%.  (c) Enrollment \textbf{D11}, evaluation \textbf{D12}, overall accuracy 93.40\%.  (d) Enrollment \textbf{D12}, evaluation \textbf{D11}, overall accuracy 96.40\%.}
	\label{fig:antenna_polarization_evaluate}
\end{figure}

Dual polarization refers to the antennas of transmitter and receiver have the orthogonal polarization directions. 
We use \textbf{D2} for enrollment and \textbf{D11} for evaluation,  which are collected at Location B and the only difference is the antenna direction. The classification result is shown in Fig.~\ref{fig:B_antenna}.
It can be seen that nearly all the packets from DUT 36 are misclassified as DUT 33. A similar result is obtained when \textbf{D6} is used for enrollment and \textbf{D12} for identification, both of which are collected at Location F. 

Linear polarization refers to the antennas of transmitter and receiver have the same polarization direction. 
We  use \textbf{D11} and \textbf{D12} for enrollment and evaluation, which are collected at different locations but with the same antenna direction. The results are given in Fig.~\ref{fig:B_G_antenna} and Fig.~\ref{fig:G_B_antenna}. Their accuracies are both above 90\% and there are no seriously misclassified DUTs. This shows the classification is not affected by the location but the antenna polarization. 
The enrollment and identification data should have the same antenna polarization for good performance. 

\section{Related work}\label{sec:related_work}

Deep learning-based RFFI usually utilizes a classification neural network to directly identify devices. It eliminates the need for feature engineering and usually leads to better performance.
However, deep learning-based RFFI still has some obstacles, two of which are system scalability and the impact of the wireless channel.

Many previous deep learning-based RFFI schemes lack scalability~\cite{shen2021radio,sankhe2019no,shen2021jsac,yu2019robust,zhang2021radio}. More specifically, it neither supports efficient device joining and leaving nor rogue device detection ability. This is because previous methods usually rely on the softmax layer for classification, and once the training was completed, the number of neurons in this layer cannot be changed.
When a new device joins and an old device leaves, the neural network should be updated by re-training, which is not practical and time-consuming. Chen~\etal use transfer learning to speed up the re-training process, but it still requires the RFFI system to have an expensive GPU. 
More seriously, such design makes the system can only identify devices that present in the training set, but rogue devices are never available during training. The rogue devices will be classified as one of the legitimate devices which is unacceptable.
To this end, the authors in~\cite{gritsenko2019finding,hanna2020open,soltani2020rf,roy2019rfal} have proposed several rogue device detection schemes. However, some require more than one neural networks to be deployed, and none of them is scalable when a new legitimate device joins the system.

Last but not least, deep learning-based RFFI is not robust to  wireless channels~\cite{al2020exposing}. 
Sankhe \etal~propose the ORACLE system to combat channel effects~\cite{sankhe2019no,sankhe2019oracle}. However, it needs to intentionally introduce impairments to the transmitter, which is costly and not suitable for IoT applications.
Data augmentation is used to combat the channel problem~\cite{soltani2020more,cekic2020robust,merchant2019enhanced,al2021deeplora,piva2021tags}, but designing an accurate channel simulator that matches the real application scenarios is challenging. This paper for the first time involves Doppler shift to emulate the real channel and evaluate its performance.

\section{Conclusion}\label{sec:conclusion}



In this paper, we propose a scalable and channel-robust RFFI framework that exploits the device-intrinsic hardware impairments for device authentication. Specifically, we leverage the deep metric learning to train an RFF extractor that has excellent generalization ability. When a new device joins the system, it only needs to send several packets for enrollment and the RFF extractor does not need to be re-trained. The k-NN algorithm is used for rogue device detection and device classification. To overcome the channel effect, we design the channel independent spectrogram and further use data augmentation to improve the system robustness to channel variations. We conduct extensive experiments using 60 commercial off-the-shelf LoRa devices. We demonstrate our framework has an excellent generalization performance for both device classification and rogue device detection. The channel independent spectrogram and data augmentation are shown to be effective under extensive tests with various channel conditions. We also find the antenna polarization affects the classification performance.


\bibliographystyle{IEEEtran}
\bibliography{IEEEabrv,mybibfile}

\end{document}